\documentclass[preprint,12pt]{elsarticle}

\usepackage{amssymb}

\usepackage[figuresright]{rotating}

\newcommand {\sqnr} {$\sqrt{s_{_{NN}}}$~=~200~GeV }
\newcommand {\ee} {$e^+e^-$ }

\def\NIM{Nucl. Instr. and Meth.}
\def\NIMA{{Nucl. Instr. and Meth.}~{\bf A}}
\def\NPA{{Nucl. Phys.}~{\bf A}}

\def\PLB{{Phys. Lett.}~{\bf B}}

\def\PRL{Phys. Rev. Lett.\ }

\def\PRC{{Phys. Rev.}~{\bf C}}

\def\EPJC{{Eur.~Phys.~J.}~{\bf C}}

\begin{document}

\begin{frontmatter}




\title{Hadron Blind Cherenkov Counters}

\author[1]{I.Tserruya}
\author[2]{K.Aoki}
\author[3]{C.Woody}

\address[1] {Weizmann Institute of Science, Rehovot ~76100, Israel}
\address[2]{KEK, High Energy Accelerator Research Organization, Tsukuba, Ibaraki 305-0801, Japan}
\address[3]{Brookhaven National Laboratory, Upton, NY ~11973 ~USA}

\begin{abstract}
A Hadron Blind Detector, or HBD, is a novel form of Cherenkov counter that is designed to provide high efficiency for detecting highly relativistic electrons while being essentially blind to most hadrons.  
In this paper, we present some of the most prominent realizations of the HBD concept in real experiments. We describe the first implementation of an HBD that was made in the CERES experiment at CERN using a spectrometer based on a doublet of hadron blind RICH detectors for the measurement of low-mass electron pairs in pA and AA collisions at the SPS. 
We next present a detailed account of a more extensive realization of the HBD that was made in the PHENIX experiment for the measurement of low-mass electron pairs in central heavy-ion collisions at RHIC, followed by a description of a very similar  detector that is currently under construction at J-PARC for the measurement of vector mesons though their \ee decay in pA collisions. We conclude with a brief discussion of possible evolutions of the HBD concept as well as possible developments and uses of HBDs in experiments at the future Electron Ion Collider.

\end{abstract}


\end{frontmatter}



\newpage
 \section{Introduction}
      \label{sec:intro}
 
Cherenkov counters are a prime choice for electron identification with high efficiency in high-energy experiments and relativistic heavy-ion physics. They generally consist of a gas radiator that emits Cherenkov photons when electrons pass through it, coupled directly, or through a window, to a detector element. The latter includes a photocathode to convert the Cherenkov photons into photoelectrons and a multiplication element like a parallel-plate avalanche counter PPAC, multi-wire proportional chamber MWPC, or gas electron multiplier GEM.

\begin{figure}[h]
 \begin{center}
  \includegraphics[width=7cm] {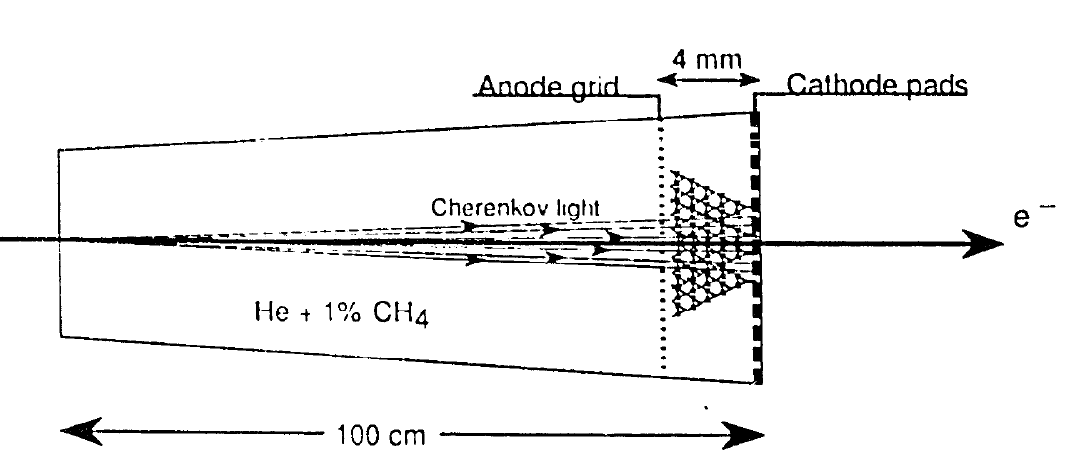}
   \caption{Schematic picture of the hadron blind detector proposed in ref. \cite{Giomataris-Charpak}.}
\label{fig:giomataris-charpak}
 \end{center}      
\end{figure}

The name Hadron Blind Detector (HBD) was coined in a paper by Giomataris and Charpak \cite{Giomataris-Charpak} where they presented the concept of a detector that would be mostly insensitive to hadrons while being highly sensitive to electrons. The principle of the detector is schematically shown in Fig.\ref{fig:giomataris-charpak}. It consists of  a gas radiator filled with He + 1\% CH$_4$ at atmospheric pressure directly coupled through a thin anode grid to a gas detector. The latter is a 4 mm gap PPAC with sub-nanosecond time resolution and with a suitable (e.g. CsI) photocathode. Electrons traversing the
radiator emit Cherenkov photons that hit the photocathode within a spread of $\sim$ 8 mm radius. Photoelectrons created on the photocathode are extracted and then amplified by the strong electric field of the PPAC and collected on the anode grid. With a 1~m long radiator, 10 photons are emitted per electron and 3 photoelectrons are expected to be released from a solid CsI photocathode adsorbed with tetrakis (dimethylamine) ethylene (TMAE). In this proposed detector concept, the hadron blindness is achieved by the high gamma threshold of the radiator gas, ensuring that only high momentum charged hadrons emit Cherenkov photons (for example, for pions, the threshold is 17 GeV/c), and by restricting the ionization charge collection to a time window of about 2 ns corresponding to a gas thickness of $\sim 200~\mu$m above the photocathode. However, this conceptual design was never implemented in an actual detector. 

The first HBDs were in fact Ring Imaging Cherenkov (RICH) counters with a gas radiator, although the term HBD is seldomly used in this context. They are hadron blind in the sense that in a gas radiator, most of the charged hadrons do not produce Cherenkov radiation, whereas essentially all electrons produce rings of asymptotic radius. The RICH technique was first described in detail in a seminal paper by Seguinot and Ypsilantis in 1977 \cite{Ypsilantis-RICH}, but it took more than a decade before it developed into a powerful experimental device. Most of the first RICH detectors were developed for particle identification (see for example refs.\cite{DELPHI, Omega, Cleo-III, DIRC}).
The CERES experiment, dedicated to the measurement of low-mass electron pairs in ultra-relativistic nuclear collisions at the CERN SPS, was the first to publish physics results using a spectrometer based on two hadron-blind RICH detectors \cite{CERES-SAu, CERES-NIM94}. The RICH detectors used CH$_4$ as radiator gas that has a high gamma threshold $\gamma_{th}$. They provided redundant electron identification with high efficiency while being blind to the copious flux of forward-going hadrons produced in fixed-target nuclear collisions.

An HBD, very different from the one proposed in \cite{Giomataris-Charpak}, was developed as an upgrade of the PHENIX detector at RHIC to enable a measurement of low-mass electron pairs in heavy-ion collisions. The original configuration of the PHENIX detector had good electron identification capabilities based on a RICH detector and an electromagnetic calorimeter. However, the set-up lacked the ability to recognize and reject the overwhelming yield of combinatorial background pairs from $\pi^0$ Dalitz decays and $\gamma$ conversions. 
The task of the HBD was to reduce these two main electron background sources.

The PHENIX HBD consisted of a Cherenkov radiator that is directly coupled in a windowless configuration to a photon detector consisting of a three-stage GEM detector with a CsI photocathode. Both the radiator and the GEMs are operated with pure CF$_4$ in a common gas volume. The detector was constructed after extensive R{\&}D to demonstrate the validity of the concept (see Refs. \cite{Kozlov, Fraenkel}  for the R{\&}D results and \cite{Tserruya-2006, Woody-IEEE-2006-2009} for other reports related to the HBD). It was then successfully operated during the 2009 and 2010 runs at RHIC for the measurement of electron pairs in p+p and Au+Au collisions, respectively \cite{dileptons-PRC2016}. A complete report on the design, construction, operation and performance of the HBD can be found in \cite{Anderson-2011}.

Another HBD was developed \cite{e16-hbd1,e16-hbd2} and is under construction in Japan for the E16 experiment at J-PARC, which aims at the measurement of electron-positron pairs produced in fixed-target p+A collisions at 30 GeV \cite{e16-tdr}. The detector has the same design as the PHENIX HBD but with a smaller hexagonal pad size. This results in a higher granularity of the pad-readout electrode allowing a cluster size analysis of the hits, thereby achieving a larger hadron rejection factor while preserving acceptable electron efficiency. 

This paper is organized as follows. Section 2 gives a short description of the CERES spectrometer. The PHENIX HBD is presented in Section 3, including the motivation for the detector, a detailed description of the detector concept, construction, operation and performance. The J-PARC HBD is presented in Section 4 including the motivation for the detector, its mechanical design and the performance of a prototype. Future possible extensions and applications of the HBD concept are described in Section 5. A short summary is given in Section 6.

 \section{The CERES spectrometer}
CERES (CErenkov Ring Electron Spectrometer) pioneered the measurement of low-mass electron pairs (m$_{e^+e^-} \leq$ 1 GeV/c$^2$) in ultra-relativistic heavy-ion collisions at the CERN SPS.
  
Dileptons are unique diagnostic tools of the quark-gluon plasma formed in these collisions. They are sensitive to the chiral symmetry restoration phase transition expected to take place together with the deconfinement phase transition and also to the thermal radiation emitted by the plasma \cite{reviews-Rapp-Tserruya}. 
However, the measurement of electron pairs is also a very challenging one. The main difficulty is the need to detect a very weak source of electron pairs, with a typical strength of the order of $\sim 10^{-5}/\pi^0$, in the presence of (i) trivial pairs from $\gamma$ conversions and $\pi^0$ Dalitz decay that are of the order of $\sim 10^{-2}/\pi^0$ and lead to a huge combinatorial background and (ii) a large yield of charged particles with a particle density $dN_{ch}/dy \sim400$ per unit of rapidity in central Pb+Pb collisions at 158 AGeV \cite{NA49}. 
A detector, highly efficient for electron detection and blind to charged hadrons, is consequently a pre-requisite for di-electron measurements.

\begin{figure}[th]
 \begin{center}
   \includegraphics[width=7.5cm] {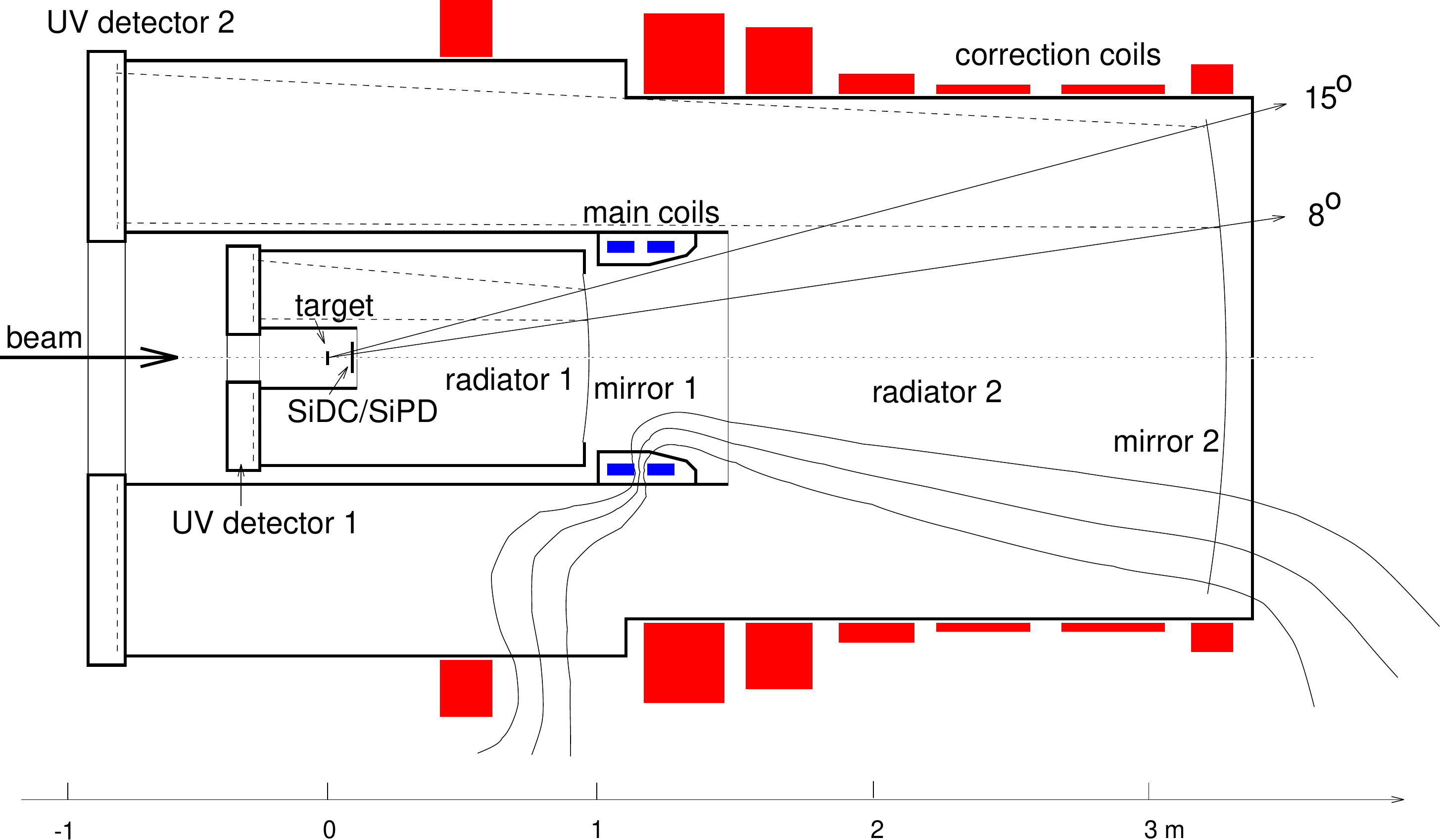}
    \caption{Layout of the original CERES double RICH spectrometer. In the lower part, magnetic field lines are shown to illustrate the field shaping \cite{CERES-NIM94}.} 
\label{fig:ceres-layout} 
 \end{center}
\end{figure}

The original configuration of the CERES spectrometer is schematically shown in Fig. \ref{fig:ceres-layout} and a detailed description can be found in \cite{CERES-NIM94}. The spectrometer covered the pseudorapidity region close to mid-rapidity $2.10 < \eta < 2.65$ with 2$\pi$ azimuthal coverage. It consisted of two coaxial azimuthally symmetric RICH detectors, one (RICH-1) located before, the other (RICH-2) behind a superconducting double solenoid, that provided electron identification and directional tracking.  
 
The radiator gas in both RICH detectors was CH$_4$ at atmospheric pressure. Cherenkov photons emitted in the radiator are reflected by a spherical mirror and focused onto a ring at the focal plane of the mirror where the UV detectors are located. The hadron blindness of the CERES spectrometer is primarily ensured by the high Cherenkov threshold $\gamma_{th}\sim$ 32 of the radiator gas such that only pions with a momentum larger than $\sim$ 4.5 GeV/c produced photons and those pions could be separated from electrons by their ring radius, up to about 15 GeV/c \cite{CERES-NIM94}. The hadron blindness in CERES was further enhanced by the location of the UV detectors, upstream of the target, such that they were not traversed by the large flux of forward-going particles produced in fixed-target nuclear collisions. 

The double solenoid provided an azimuthal deflection of charged particles for momentun determination leaving the polar angle unchanged. The current in the two solenoids was adjusted such that the magnetic field was compensated to nearly zero in the region of the RICH-1 radiator thus preserving the original direction of the particles. This feature is essential to enable identification of pairs from $\gamma$ conversions and $\pi^0$ Dalitz decays which typically have very small opening angles (close pairs). In the region of the RICH-2 radiator, the field was shaped by a set of correction coils to point back to the vertex, ensuring straight particle trajectories and therefore sharp ring images (see the field lines in the lower part of Fig. \ref{fig:ceres-layout}).
 
The set-up also included a novel silicon drift chamber (SiDC) that provided high-resolution vertex reconstruction and particle tracking to the interaction point \cite{silicon-drift-NIM93} and a 64-pad silicon detector that was used to characterize the event centrality and to provide the first level trigger \cite{gunzel-NIM92}.

The unconventional design of the CERES spectrometer that provided directional tracking but no real particle tracking proved to work very well in low-multiplicity collisions like in p+Be or p+Au \cite{CERES-pBe-pAu}, reaching a pair reconstruction efficiency of 50\% with a signal to background ratio (S/B) of 1/2. However, as the multiplicity increases, the performance deteriorates. The UV detector occupancy increases leading to fake rings and to a decreased reconstruction efficiency. It was still acceptable for central S+Au collisions 
where the efficiency was 9\%  with S/B $\sim$ 1/6 \cite{CERES-SAu}. However, it was evident that the concept would fail at the higher particle densities that were expected in central Pb+Pb collisions. For the measurements with Pb beams \cite{CERES-PLB98-EPJC2005, CERES-PLB2008} the set-up had to be supplemented with additional detectors that provided particle tracking \cite{CERES-NIM96} and helped the pattern recognition in the two RICHes. A doublet of Si drift chambers, located $\sim$ 10 cm downstream of the target, replaced the previously used single SiDC to enable charged particle tracking into RICH-1 and to reject close tracks \cite{doublet-SiDC-NIM96}. A pad chamber (MWPC with pad readout) was installed immediately behind the mirror of RICH-2 to provide tracking downstream of RICH-2. Later, in order to improve the mass resolution, the pad chamber was replaced by a cylindrical TPC with a radial electric field \cite{CERES-TPC2008}.

In all the di-electron measurements performed by CERES, the hadron-blind double RICH spectrometer performed very well. On the technical side, RICH-1 achieved an effective figure of merit of $N_0$ (see Section 3.4 for a definition of $N_0$) of 131 cm$^{-1}$ \cite{CERES-NIM94}, a record for a gas Cherenkov detector at the time, to be surpassed only much later by the PHENIX HBD described in the next section. On the physics side, a strong enhancement of low-mass electron pairs, possibly connected to chiral symmetry restoration, was observed in all the nucleus-nucleus collisions studied by CERES \cite{CERES-SAu, CERES-PLB98-EPJC2005, CERES-PLB2008, CERES-PRL2003}.

\section{The PHENIX Hadron Blind Detector}
\label{}

\subsection{Motivation}
The PHENIX HBD was developed as an upgrade of the PHENIX detector at RHIC for the measurement of low-mass electron-positron pairs produced in central Au+Au collisions at energies up to \sqnr. 

The challenge of the measurement is the same as described in the previous section for the CERES experiment and it is even enhanced by the higher multiplicities of charged particles, which are larger by about a factor of 2 at RHIC compared to the SPS \cite{PHENIX-2001}.
 
The original configuration of the PHENIX detector had good electron identification capabilities based on a RICH detector and an electromagnetic calorimeter. However, the set-up lacked the ability to recognize and reject the overwhelming yield of combinatorial background pairs originating from $\pi^0$ Dalitz decays and $\gamma$ conversions. 
For example, in the invariant mass range $m \sim$ 0.15 - 0.75 GeV/$c^2$, the signal to background ratio in minimum bias Au+Au collisions, where the signal is estimated by the cocktail of known sources, was S/B $\simeq$ 1/600 \cite{dileptons-PRC2016}. Under these conditions, an uncertainty of 1\% in the background subtraction, induces a change of a factor 6 in the signal, making the measurement of the low-mass pair continuum very challenging. 

In order to perform this measurement in central Au+Au collisions at RHIC energies, a new detector was required that could provide additional electron identification, suppress the combinatorial background and improve the signal sensitivity.
The detector had to achieve an electron identification efficiency of $\sim$ 90\% and provide a hadron rejection factor of $\sim$ 50. This was accomplished using the HBD, which was a highly efficient Cherenkov detector that was designed to have minimum sensitivity to charged particles other than electrons \cite{Anderson-2011}. 

\begin{figure}[th]
 \begin{center}
   \includegraphics[width=7.5cm] {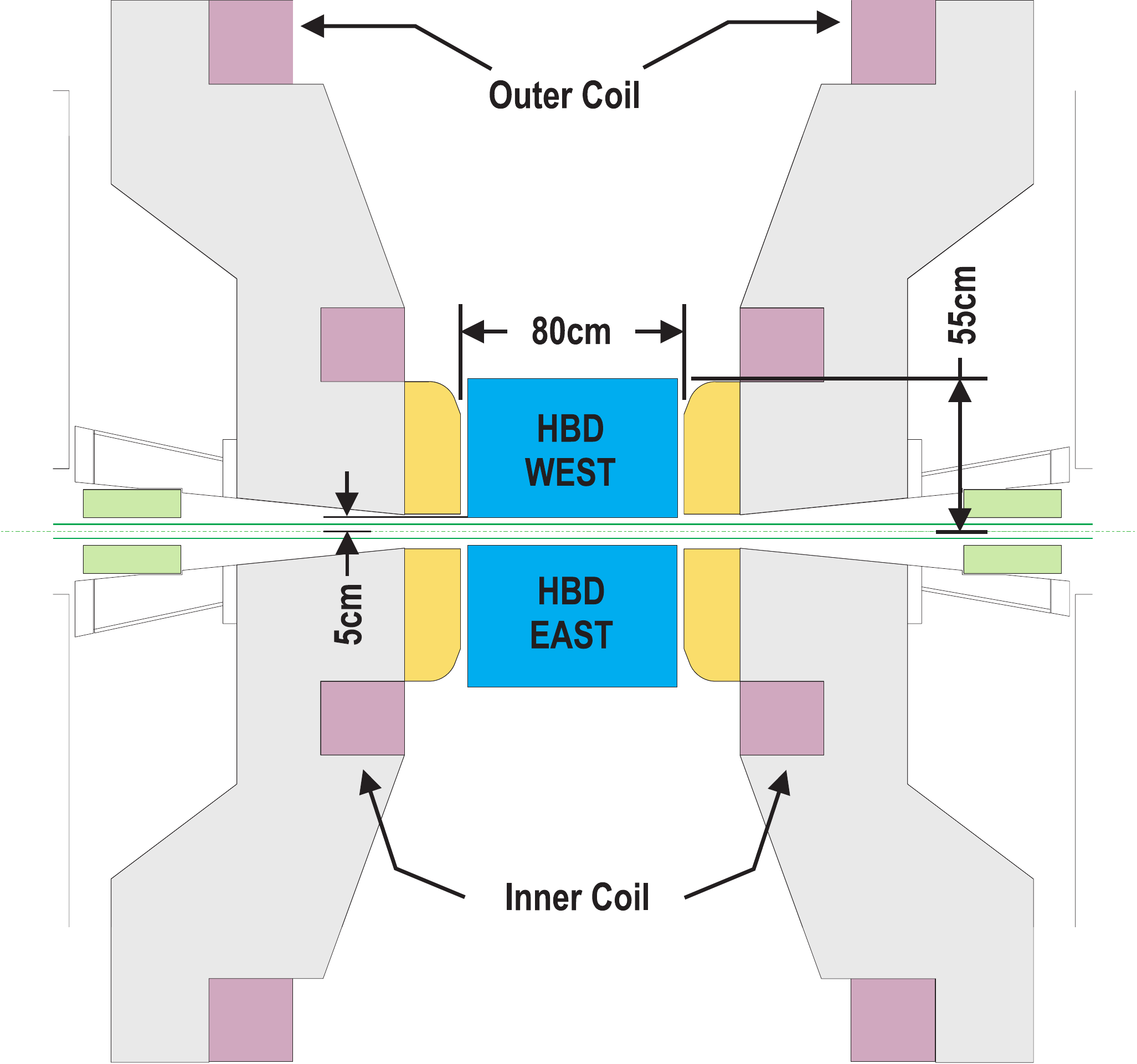}
   	\caption{\label{fig:PHENIX_Detector_with_HBD} {The PHENIX Detector at RHIC with the HBD located inside the inner coil of the central magnet \cite{Anderson-2011}.} }
 \end{center}
\end{figure}

As shown in  Fig.~\ref{fig:PHENIX_Detector_with_HBD}, the HBD was located in a nearly field-free region of the PHENIX spectrometer that was created by two Helmholtz coil pairs (inner and outer) with opposing fields. This allowed electron pairs from Dalitz decays and conversions, which are produced with a very small opening angle, to pass through the detector together without being affected by the magnetic field. Electron pairs with a larger invariant mass, which constitute the signal, are produced with a larger opening angle and pass through the detector as two well separated tracks. It was therefore possible to reject background pairs by requiring the amplitude of the Cherenkov signal produced in the HBD to be consistent with that of a single electron whereas those with twice the amplitude or those producing two close hits were rejected. The HBD was also designed to have larger coverage in azimuthal angle $\phi$ and pseudorapidity $\eta$ than the main PHENIX Central Arm detectors in order to detect electrons outside the PHENIX acceptance that could produce unpaired electrons inside the fiducial region and contribute to the low-mass pair background. 
  
\subsection{Detector Concept}
   The PHENIX HBD is a windowless proximity focused Cherenkov counter that is comprised of a gas radiator and a photon detector consisting of a three-stage GEM detector with a CsI photocathode \cite{Kozlov, Fraenkel}. The radiator and photon detector use CF$_4$ in a common gas volume such that there is no window separating the two which allows for maximum yield and efficiency for transmitting UV light from the radiator onto the CsI photocathode. The gas CF$_4$ is known to scintillate. It was nevertheless chosen as the radiator gas because of its very large bandwidth from $\sim$ 6 eV given by the threshold of CsI to $\sim$ 11.1 eV given by the 50\% CF$_4$ cut-off (zero at $\sim$ 12.4 eV). This resulted in a very high figure of merit $N_0$ with a calculated value of 714 cm$^{-1}$ under ideal conditions and to a very high photon yield (see Section 3.4 below). However, a very high level of gas purity was required since even trace amounts of oxygen and water could produce absorption in the gas that could lead to reduced photon detection efficiency. Another advantage of CF$_4$ is that it can also serve as operating gas for the triple GEM detector, a necessary condition for the windowless configuration. However, for a gain of 3$\times$10$^3$ for example, CF$_4$ requires about 150 V more high voltage across the GEMs compared to a more conventional gas like 70/30  Ar/CO$_2$ \cite{Kozlov}. This imposed strict clean conditions during the GEM assembly to avoid any dust that would prevent reaching the operating voltage.

   Since there is no mirror or lens to form a ring, Cherenkov photons from relativistic particles passing through the radiator are emitted in a cone along the particle trajectory and form a "blob" with an approximately circular area on the photocathode plane. The nominal length of the radiator is 50 cm and the maximum size of the blob is $\sim$ 9.9 cm$^2$. This allows for a rather coarse segmentation of the readout plane, which is divided into hexagonal pads measuring 1.55 cm on a side and have an area of 6.2 cm$^2$. This serves to reduce the channel count for the readout electronics and hence the overall cost. However, the pad area is small enough that most Cherenkov blobs spread over more than one pad, while any hadron signal would be contained within a single pad.
   
\begin{figure}[th]
 \begin{center}
   \includegraphics[width=7.5cm] {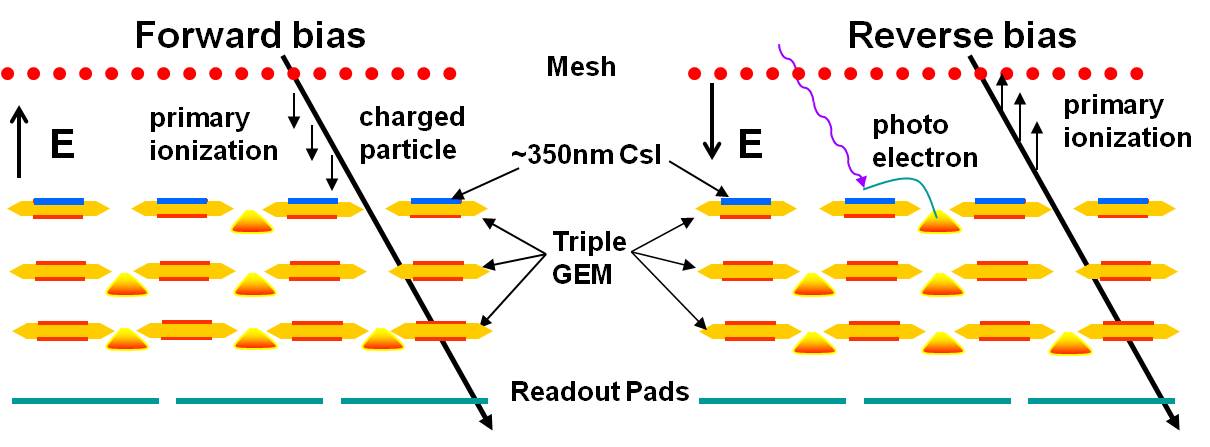}
   	\caption{\label{fig:GEM_stack_FB_RB_concept} {Triple GEM stack operated in Forward Bias mode (left) where the primary ionization produced in the gap between the mesh and upper GEM foil is collected by the GEM detector, and Reverse Bias mode (right) where the ionization produced in the gap is repelled away from the GEM detector \cite{Anderson-2011}.} }
 \end{center}
\end{figure}

   The hadron blindness feature of the detector was achieved by introducing a mesh electrode just above the CsI photocathdode evaporated on the upper GEM foil (as shown in Fig.~\ref{fig:GEM_stack_FB_RB_concept}) that could be biased to produce two different modes of operation. In Forward Bias mode, the mesh was biased to force electrons produced by ionization in the gap between the mesh and the upper GEM towards the GEM detector which would then be collected and amplified, making the detector sensitive to all charged particles. In Reverse Bias mode, the mesh was biased to force electrons produced by ionization in the gap away from the GEM detector, making it essentially insensitive to charged particles and only sensitive to electrons produced on the photocathode by UV light. Only ionization in a very small region ($\sim$ 100 $\mu$m) above the top GEM foil was collected by the GEM detector and contributed to the resulting Cherenkov signal. Figure~\ref{fig:GEM_stack_FB_RB_concept} shows the two different modes of operation that illustrate this concept.
   
\subsection{Detector Construction}

\begin{figure}[th]
 \begin{center}
   \includegraphics[width=7.5cm] {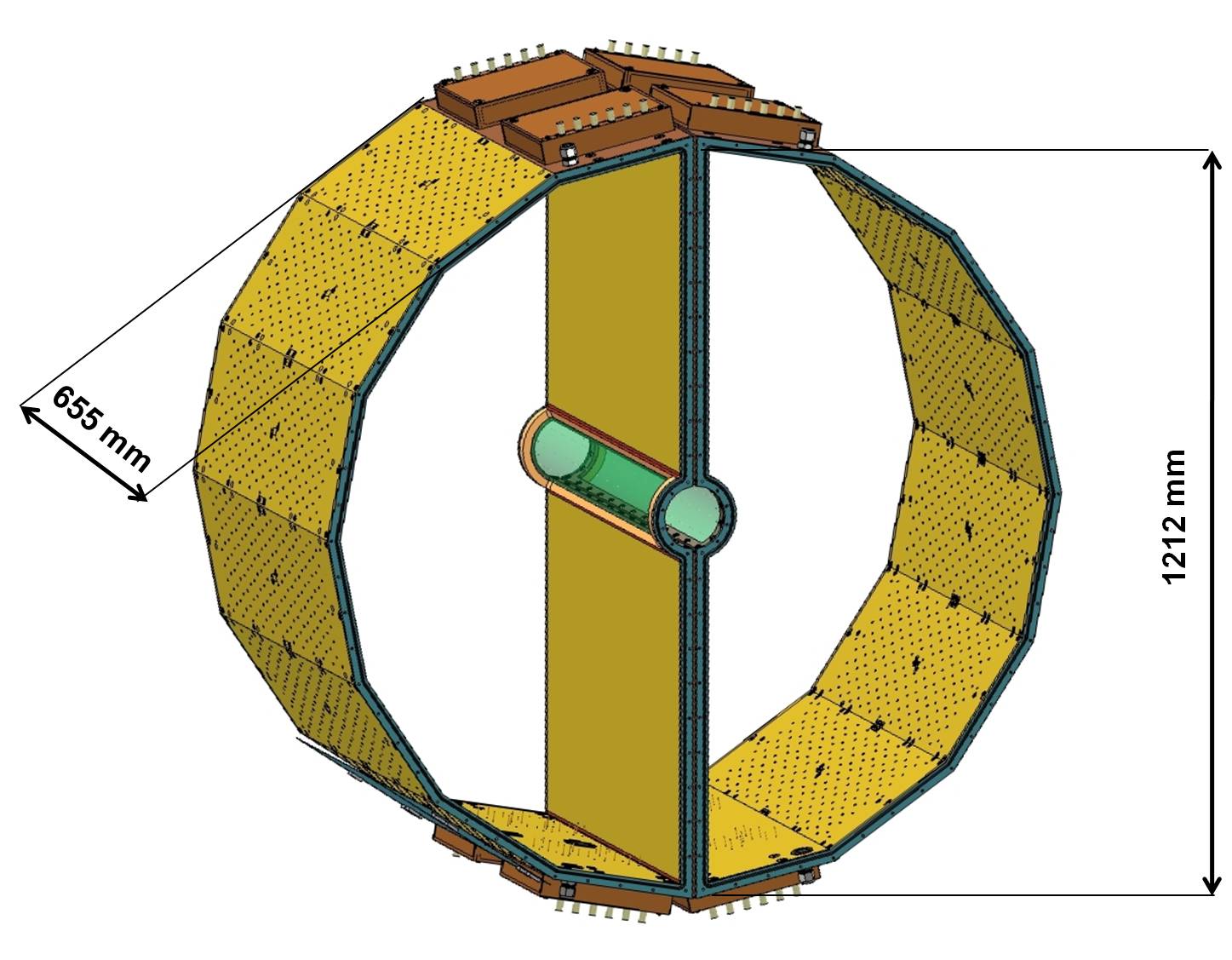}
   	\caption{\label{fig:3D_View} {3D view of both arms of the HBD \cite{Anderson-2011}.} }
 \end{center}
\end{figure}

  The HBD consisted of two identical arms located on the east and west sides of the PHENIX spectrometer as shown in Fig.~\ref{fig:PHENIX_Detector_with_HBD}. It extended from a radius of $r \sim 5$ cm near the beam pipe to $r \sim 60$ cm at the outer radius and had a total length of $\sim$ 130 cm along the beam direction. Fig.~\ref{fig:3D_View} shows a 3D view of both arms. 
Each arm contained 10 modules of triple GEM detectors and their readout electronics, covering 112.5$^o$ in $\phi$ and $\pm$0.45 units in $\eta$. This was considerably larger than the 90$^o$ in $\phi$ and $\pm$0.35 units in $\eta$ covered by the PHENIX central arm detectors and provided an effective veto area for rejecting background electrons.        


 \begin{figure}[th]
 \begin{center}
   \includegraphics[width=7.5cm] {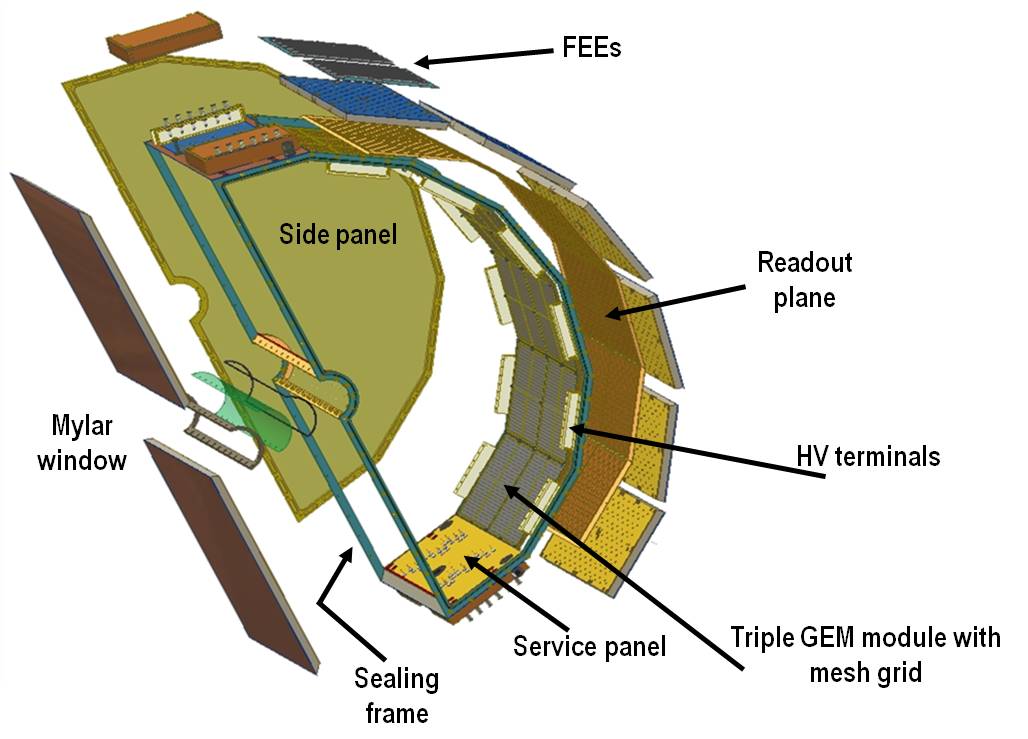}
   	\caption{\label{fig:Arm_Exploded_View} {Exploded view of one of the HBD arms \cite{Anderson-2011}.} }
 \end{center}
\end{figure}
 
\begin{figure}[th]
 \begin{center}
   \includegraphics[width=7.5cm] {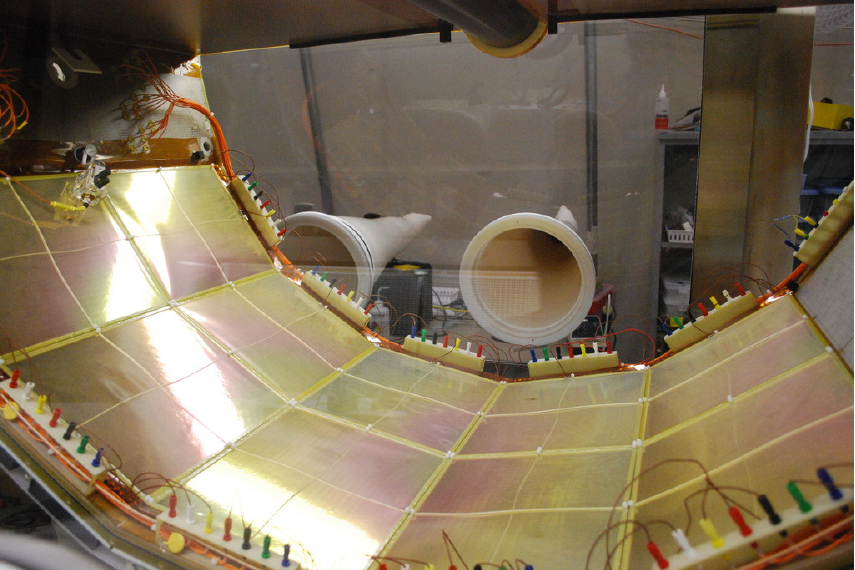}
   	\caption{\label{fig:Completed_GEM_Modules} {Completed HBD arm with CsI coated GEM modules inside the glove box   \cite{Anderson-2011} .} }
 \end{center}
\end{figure}

  Figure~\ref{fig:Arm_Exploded_View} shows an exploded view of one of the arms depicting the details of its internal construction. The GEM modules measured 23 $\times$ 27 cm$^2$ and the CsI photocathode was 
  produced in a large evaporator that deposited $\sim$ 300 nm of CsI on the upper surface. The GEM was then transferred to a glove box where it was installed in the detector with the other GEMs. Figure~\ref{fig:Completed_GEM_Modules} shows one of the completed HBD arms inside the glove box with all GEM modules installed with their CsI photocathodes. The signals from the hexagonal pads on the readout plane were connected through the outer wall of the detector (see Fig.~\ref{fig:Arm_Exploded_View}) to the front end electronics located on the outside. 
 
\subsection{Operation and Performance}

  Due to its windowless construction, the HBD achieved a high yield and a high level of efficiency for detecting UV photons from Cherenkov light which allowed the GEM detectors to be operated with relatively low gain. This was also made possible by the low-noise electronics that was used to read out the GEM signals. The GEMs were operated with a gas gain of $\sim$ 4000, which required a voltage across each GEM foil of $\sim$ 470 V. This was well below the breakdown voltage in pure CF$_4$ ($\sim$ 550 V) and allowed for stable operation. The preamp had an RMS noise of $\sim$ 1000 electrons which corresponded to a signal of $\sim$ 0.2 photoelectrons. This enabled a measurement of single photoelectrons from scintillation light produced by charged particles in the gas which uniformly illuminated the photocathode surfaces and was used to measure and monitor the gain of the entire detector on a pad by pad basis during normal operation.
  
  The overall photon detection efficiency of the HBD is determined by a number of factors. It can be quantified by the figure of merit N$_0$ that is typically used to characterize Cherenkov counters. For an ideal detector, N$_0$ would be given by the integral of the CsI quantum efficiency $QE(E)$ convoluted with the transmission of the radiator gas $T_G$ over the wavelength range of sensitivity, which for the HBD is given by:

\hfil\break
\begin{equation}
     N_0^{ideal} = 370 \int_{6.2}^{12.4}QE(E) \cdot T_G \cdot dE = 714~cm^{-1}
\\
\end{equation}

   For an actual detector, a number of other efficiency factors serve to reduce the photon yield. These include the optical transparency of the various electrodes, the collection efficiency of the photoelectrons, and the threshold for measuring a signal from the pads. A list of the various efficiency factors contributing to the HBD photon yield is listed in Table ~\ref{tab:npe} (see ref. \cite{Anderson-2011} for details). The calculated $N_0$ for the HBD taking into account all efficiencies is 328 $\pm$ 46 cm$^{-1}$. This leads to an expected number of photoelectrons N$_{pe}$ for a single electron track of N$_{pe}$= N$_0$L/$\gamma_{th}^2$ = 20.4 $\pm$ 2.9, where L is the radiator length and $\gamma_{th}$ is the average Cherenkov threshold over the sensitive wavelength range of the detector.

 \begin{table}[!h]
\caption{Figure of merit and Cherenkov photon yield \cite{Anderson-2011}.}
\label{tab:npe}
\begin{tabular}[]{l c}  \\
N$_0$ ideal value                     &  714 cm$^{-1}$   \\
\hline
Optical transparency of mesh          &  88.5\% \\
Optical transparency of photocathode  &  81\%   \\
Radiator gas transparency             &  89\%   \\
Transport efficiency                  &  80\%   \\
Reverse bias and pad threshold        &  90\%   \\
\hline
N$_0$ calculated value                & 328$\pm$ 46 cm$^{-1}$    \\
N$_{pe}$ expected                     &  20.4$\pm$ 2.9   \\
\hline
N$_{pe}$ measured                     & 20   \\
N$_0$ measured value                  &  322 cm$^{-1}$ \\
\hline
\end{tabular}
\end{table}

\begin{figure}[!htbp]
 \begin{center}
   \includegraphics[width=7.5cm] {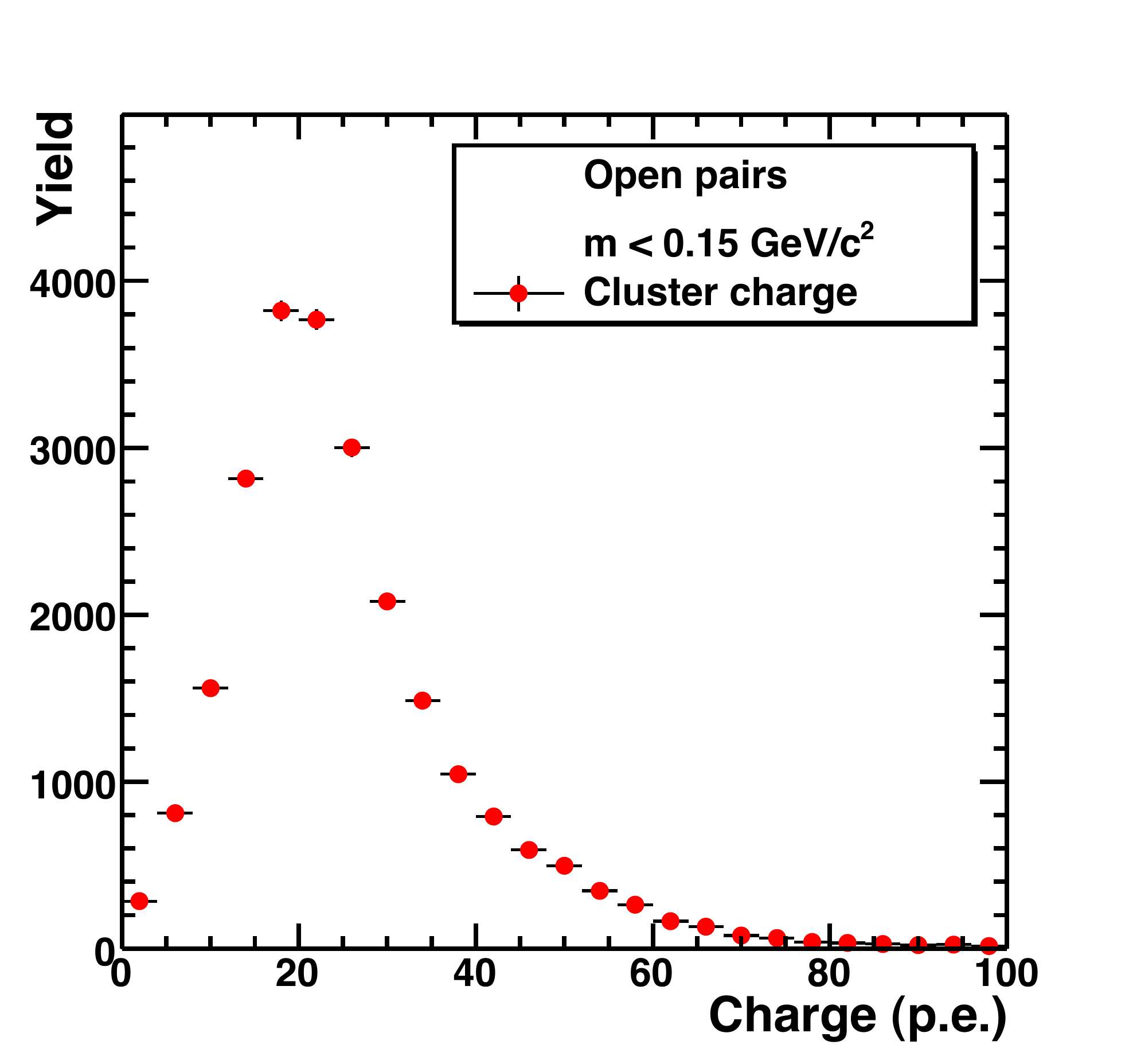}
   	\caption{\label{fig:Single_Electron_Npe} {Photoelectron yield for single electron tracks measured in the PHENIX spectrometer passing through the HBD \cite{Anderson-2011}.   } }
 \end{center}
\end{figure}

\begin{figure}[!htbp]
 \begin{center}
   \includegraphics[width=7.5cm] {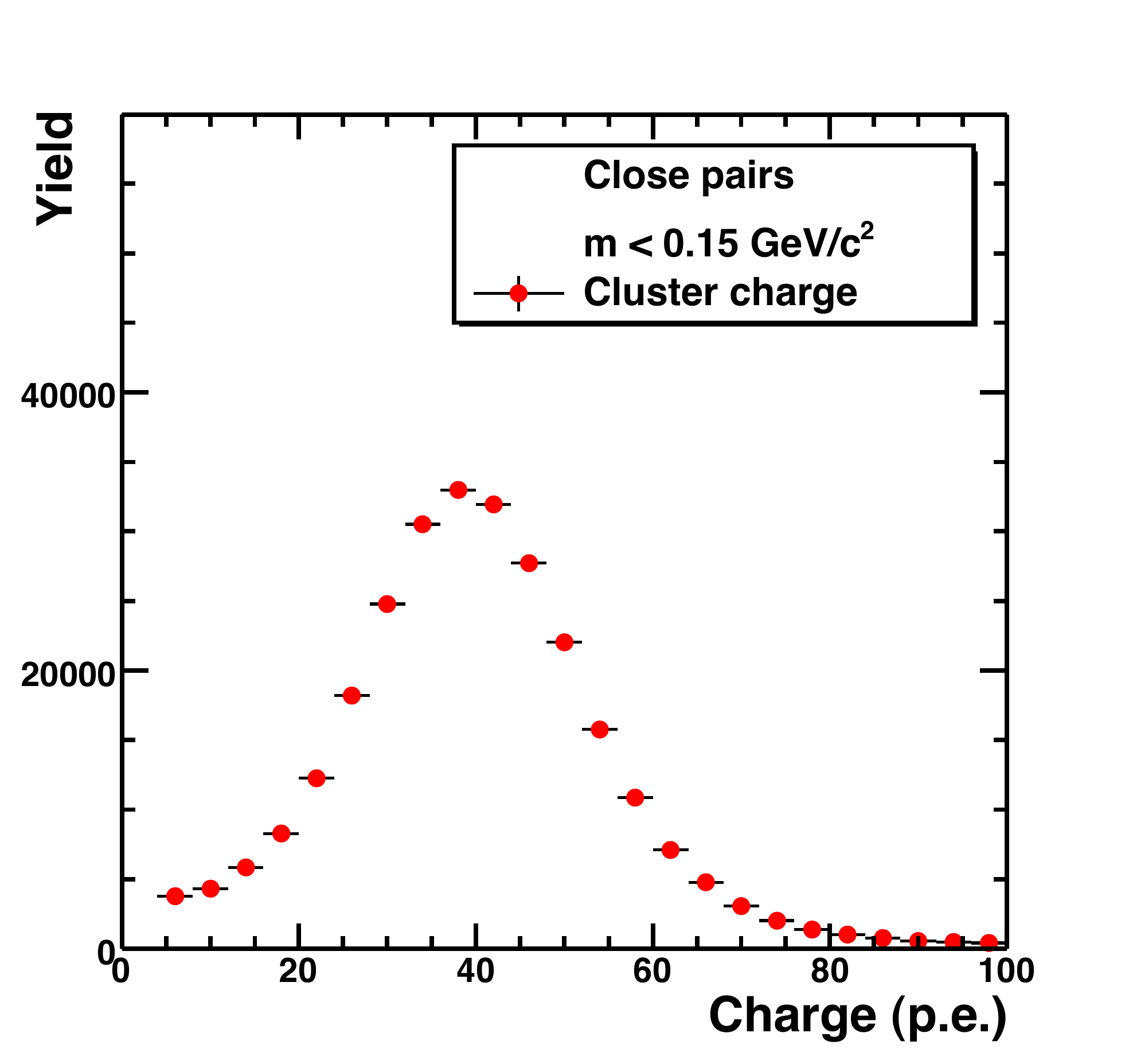}
   	\caption{\label{fig:Double_Electron_Npe} {Photoelectron yield for double electron tracks (unseparated due to lack of a magnetic field) measured in the PHENIX spectrometer passing through the HBD \cite{Anderson-2011}.} }
 \end{center}
\end{figure}

 Figure ~\ref{fig:Single_Electron_Npe} shows the spectrum of detected photoelectrons for single electron tracks measured in the PHENIX spectrometer that passed through the HBD. The mean number of photoelectrons is 20, which corresponds to an $N_0$ of 322 cm$^{-1}$, in good agreement with the expected value.  Figure ~\ref{fig:Double_Electron_Npe} shows the spectrum of photoelectrons for double electrons, which are \ee pairs that remain unseparated as they pass through the field free region of the HBD. The average number of photoelectrons for these tracks is 40, which allows an excellent separation of single and double electron tracks for rejecting Dalitz pairs and conversions.
  
\begin{figure}[!htbp]
 \begin{center}
   \includegraphics[width=7.5cm] {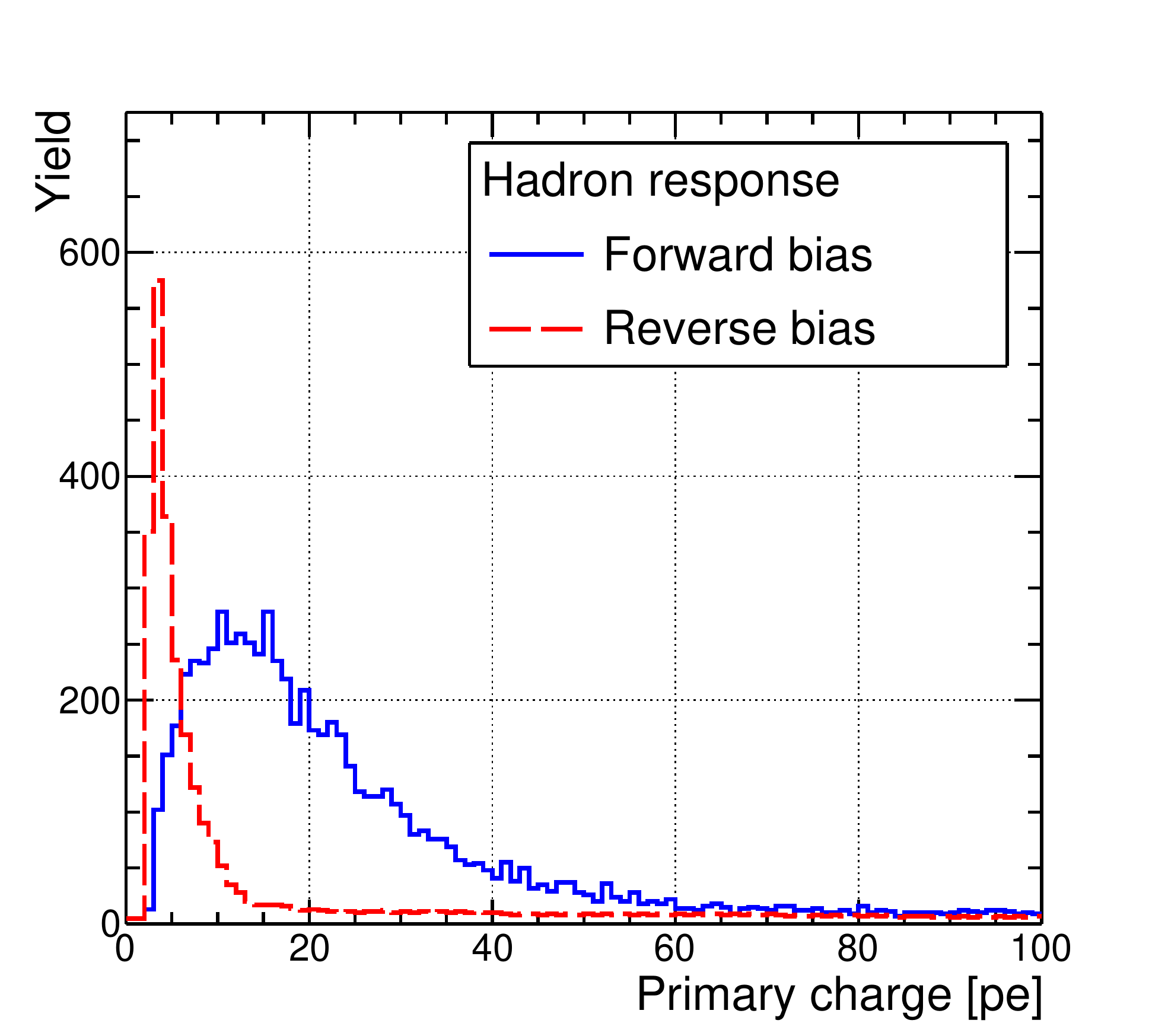}
   	\caption{\label{fig:Hadrons_FB_RB} {Pulse height spectra in the HBD for hadrons in Forward Bias (blue solid)  and Reverse Bias (red dashed) operating modes  \cite{Anderson-2011}.} }
 \end{center}
\end{figure}

\begin{figure}[!htbp]
 \begin{center}
   \includegraphics[width=7.5cm] {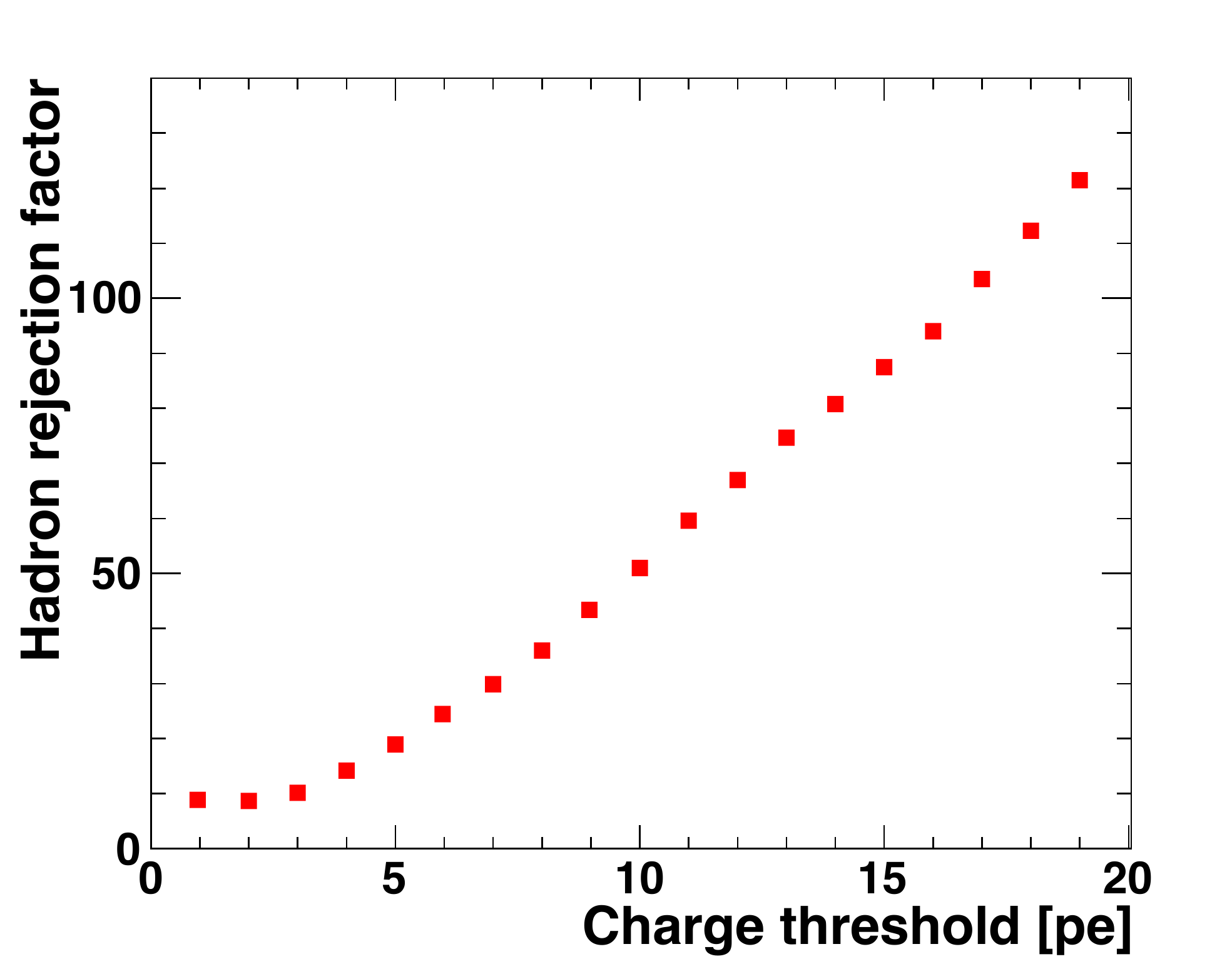}
   	\caption{\label{fig:Hadron_Rejection_Factor} {Hadron rejection factor as a function of the cut on the number of photoelectrons  \cite{Anderson-2011}.} }
 \end{center}
\end{figure}

 Figure ~\ref{fig:Hadrons_FB_RB} shows the pulse height spectrum for hadrons passing through the HBD in Forward Bias and Reverse Bias operating modes. In Forward Bias, the spectrum shows a Landau distribution from the dE/dx energy loss in the gap between the mesh and the upper GEM electronde. In Reverse Bias mode, the spectrum is highly suppressed due to the charge being directed away from the GEM detector with only a small contribution from the small region above the photocathode. Figure ~\ref{fig:Hadron_Rejection_Factor} shows the hadron rejection factor achieved as a function of the cut on the number of photoelectrons. For a cut of 10 p.e., the hadron rejection factor is $\sim$ 50, which when applied to the electron pulse height spectrum shown in  Figure~\ref{fig:Single_Electron_Npe}, results in an electron efficiency of $\sim$ 90\%.

  The PHENIX HBD was used to collect data on low-mass pairs during the 2009 and 2010 runs at RHIC. The 2010 run provided data on Au+Au collisions while the 2009 run provided data on p+p collisions that was used as the baseline for comparison to the Au+Au data. The detector performed extremely well during both runs and produced one of the most successful measurements of low-mass electron pairs in heavy-ion collisions. 
An enhancement of \ee in the low-mass region was observed, consistent with the STAR results \cite{STAR-2015} and superseding the previous PHENIX results that showed a much stronger enhancement \cite{PHENIX-ee-2010}. These results are now published and further details can be found in \cite{dileptons-PRC2016}.

\section{The Hadron Blind Detector at J-PARC}
\label{}
\subsection{Motivation}

The J-PARC E16 experiment is currently under construction to measure $e^+ e^-$ pairs from vector mesons
in p + A collisions at 30~GeV. The physics motivation is to confirm the
previously observed in-medium spectral change of vector mesons by the KEK PS E325 experiment~\cite{e325-phi, e325-rhoomega},
and to collect 100 times more statistics to conduct a systematic study.
Therefore, high statistics is one of the key goals of the experiment.
The experimental targets are very thin (0.5\% radiation length) in order to avoid
unwanted gamma conversions and bremsstrahlung which cause distortion of the $e^+e^-$ spectrum.  
Thus, the E16 experiment is designed to accept as high as 10$^{10}$ particles per 2~sec spill of the 30~GeV proton beam
and to have large acceptance to achieve its goal.
E16 has chosen an HBD for the first stage electron identification in the experiment
as its mirrorless and windowless features are suitable for providing a large acceptance spectrometer
within the space limitations.


 \begin{figure}[htbp]
 \begin{center}
   \includegraphics[width=7cm] {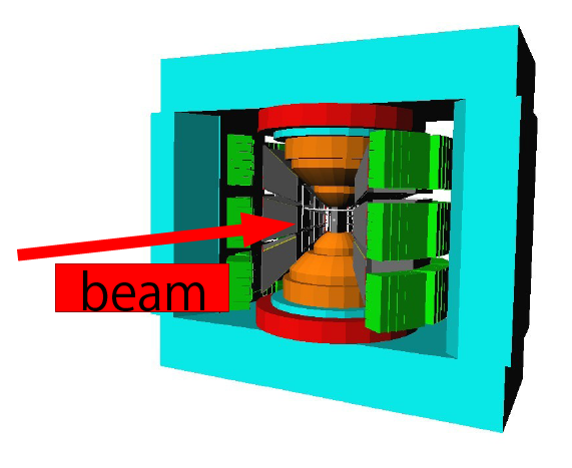}
   	\caption{\label{fig:e16-3d} {3D view of the J-PARC E16 spectrometer. 
   	Almost all the spectrometer components are installed inside a large dipole magnet.
   	} }
 \end{center}
\end{figure}

Figure \ref{fig:e16-3d} is a 3D view of the J-PARC E16 spectrometer.
The spectrometer components are placed inside a dipole magnet
and have a horizontal (vertical) acceptance of 15-135 ($\pm$45) degrees.
The detectors are divided vertically into three groups. The middle ones, which vertically cover $\pm 15$~degrees, will be prepared at the first stage of construction.

\begin{figure}[htbp]
 \begin{center}
   \includegraphics[width=7.5cm] {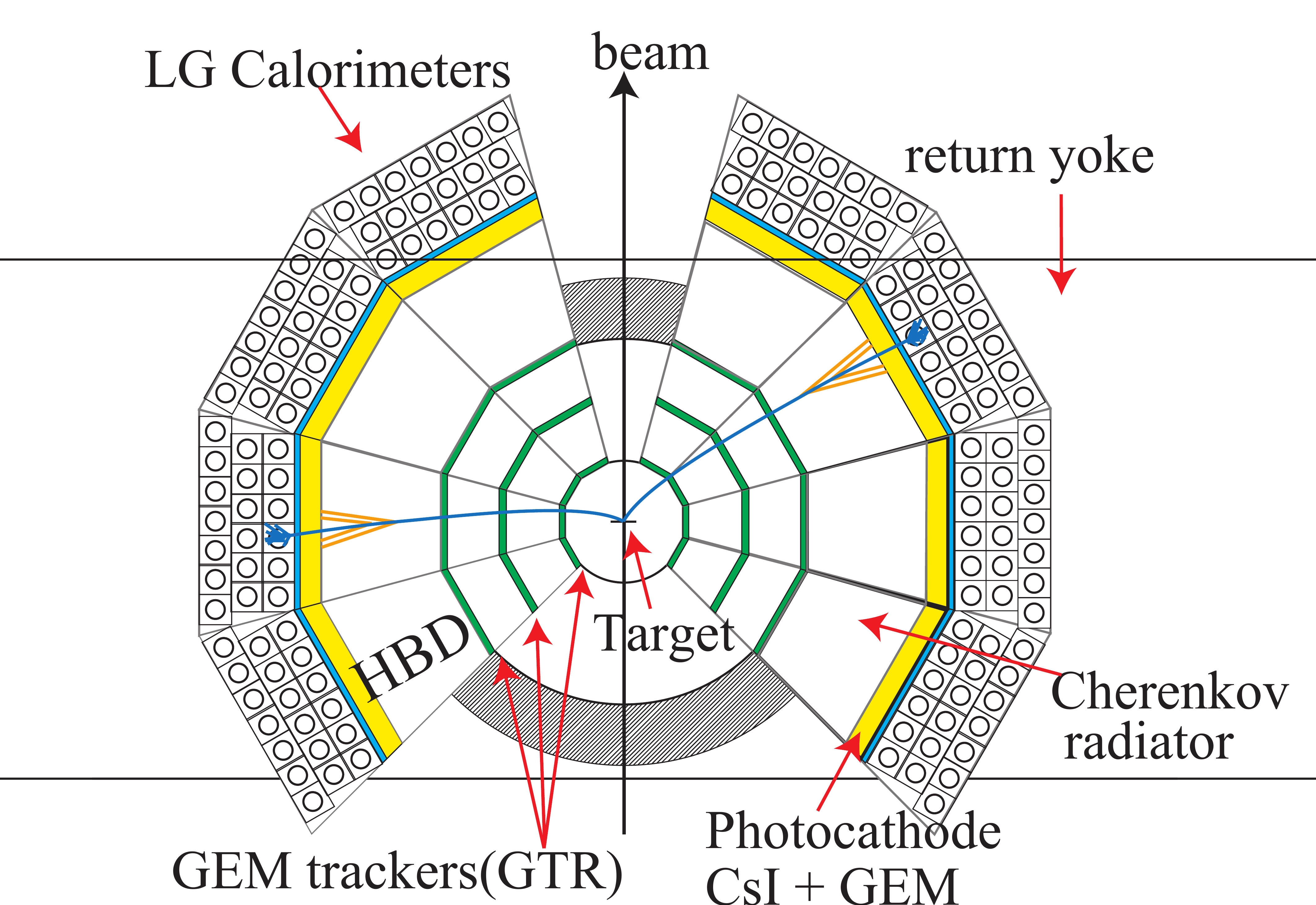}
   	\caption{\label{fig:e16-spectrometer} {Cut view of the J-PARC E16 spectrometer.
   	The experimental targets are placed at the center and are covered by
   	two tracking detectors (SSD, not visible in the figure, and GTR)
   	and two electron identification detectors (HBD, LG Calorimeters.)} }
 \end{center}
\end{figure}

Figure~\ref{fig:e16-spectrometer} displays a cut view of the spectrometer. 
At the center are the experimental targets. They are surrounded 
by a Silicon Strip Detector (SSD, not visible in the figure), three layers of GEM Trackers (GTR), the 
HBD and a lead glass calorimeter (LG).
The HBD extends from r = 690~mm to r=1280~mm.
One set of detector components forms a module and eight modules can be
seen in the figure.
Two modules of the HBD are realized in a single chamber to reduce dead areas.
The outer sides of the HBD chambers are covered with photosensitive GEMs and
the Cherenkov photons are detected directly without reflections.
The radiator is 500~mm long and uses pure CF${}_4$ at atmospheric pressure, which has a Cherenkov threshold (averaged over the sensitive bandwidth of the detector) of $\gamma_{th}$ = 28.8. This corresponds to an effective momentum threshold for pions $\sim$ 4.0~GeV/$c$ (similar to the PHENIX HBD  \cite{dileptons-PRC2016}), which is very suitable for this experiment since the momentum range of interest is 0.4-3.0~GeV/$c$.


Due to budgetary constraints, a staged approach has been adopted for the E16 experiment.
RUN0 is mainly dedicated for the commissioning of the beam line and the detector components. RUN1 is intended to be the first physics run with one third of the detector modules instrumented. 
The RUN1 configuration will have a vertical coverage of $\pm 15$~degrees.
RUN2 will be performed with full coverage as in Fig.~\ref{fig:e16-3d}
to collect high statistics.
The beam line construction is scheduled to be completed in January 2020
and RUN0 is expected to follow. RUN1 is expected in 2022 and will be after a planned long shutdown of the J-PARC Main Ring.



\subsection{Mechanical Design}
\begin{figure}[htbp]
 \begin{center}
   \includegraphics[width=5cm] {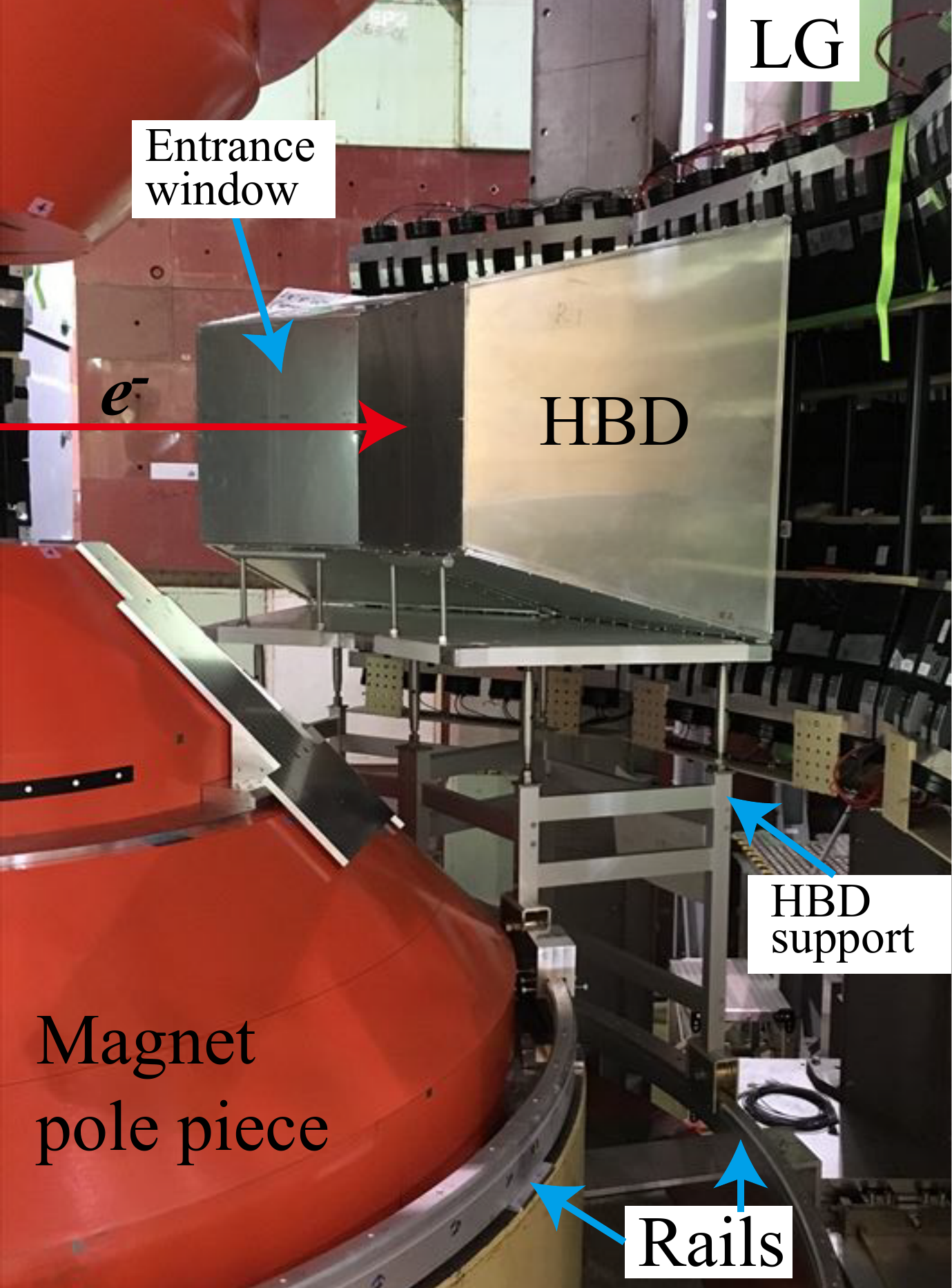}
   	\caption{\label{fig:hbd-chamber} {Photo of an HBD chamber installed for a mechanical test. The detector sits on a support which is mounted on rails. 
   	   The entrance window seen in the photo is an aluminum plate, but in the final design it will be aluminized mylar.
       } }
 \end{center}
\end{figure}

The working principle of the J-PARC HBD is the same as the PHENIX HBD.
Figure \ref{fig:hbd-chamber} shows a picture of the HBD detector installed in the spectrometer magnet for a mechanical test.
The chamber frames and the side walls are made of aluminum and welded to each other. The top and bottom sides of the chamber
 are made of aluminum and form lids that are screwed onto the chamber along with o-ring seals. The readout plane consists of hexagonal  pads which are 10~mm on a side and are realized in a conventional PCB board.
Two of the boards are glued to the right sides of the chamber. The entrance windows of the final design are aluminized-mylar and are glued on the chamber frame. The windows in the picture are aluminum plates which were installed only for test purposes. 

Electrons enter from the left side as indicated by the arrow, emitting Cherenkov light inside the radiator volume.
The inner surface on the right side of the chamber contains the photosensitive GEM detector.
The GEM detector configuration is similar as in Fig.~\ref{fig:GEM_stack_FB_RB_concept}.
The GEMs used for the E16 experiment are manufactured using a wet-etching technique by RAYTECH\footnote{http://www.raytech-inc.co.jp/}.
The GEM holes are double-conical in shape due to the wet-etching process.
The hole diameter in the copper is 55~$\mu$m, the hole diameter in the kapton at its narrowest point is 25~$\mu$m, and the hole pitch is 110~$\mu$m. We refer to this as a 55/110-GEM.
The triple GEM configuration achieves a gain of 20000 with 500 volts across each GEM.
The drift gap is 4~mm, the first and second transfer gaps are 1.0~mm and 1.5~mm respectively, and the
induction gap is 2~mm.
The size of the GEM is 295.5 $\times$ 295.5 mm${}^2$.
The top GEM is gold plated on which CsI is evaporated by Hamamatsu photonics.
The bottom side of the bottom GEM is divided into nine segments to provide a trigger signal. The HBD is placed on a support which is mounted on rails
that are used to slide the detector into the desired position.


%


\subsection{Performance of prototype HBD} 
The final design was determined based on beam tests and laboratory tests of several different prototypes. 
Here the performance of one of the prototypes~\cite{e16-hbd2} is discussed which 
demonstrates the power of the cluster size analysis. 
The differences between the prototype and the final design which may affect the performance are the following: the top GEM of the prototype was a 55/240-GEM instead of a 55/110-GEM, the readout electrodes are $10 \times 10$~mm${}^2$ square pads
instead of hexagonal pads with a side of 10~mm, and the first transfer gap is 1.5~mm instead of 1.0~mm.

 \begin{figure}[htb]
 \begin{center}
   \includegraphics[width=6cm] {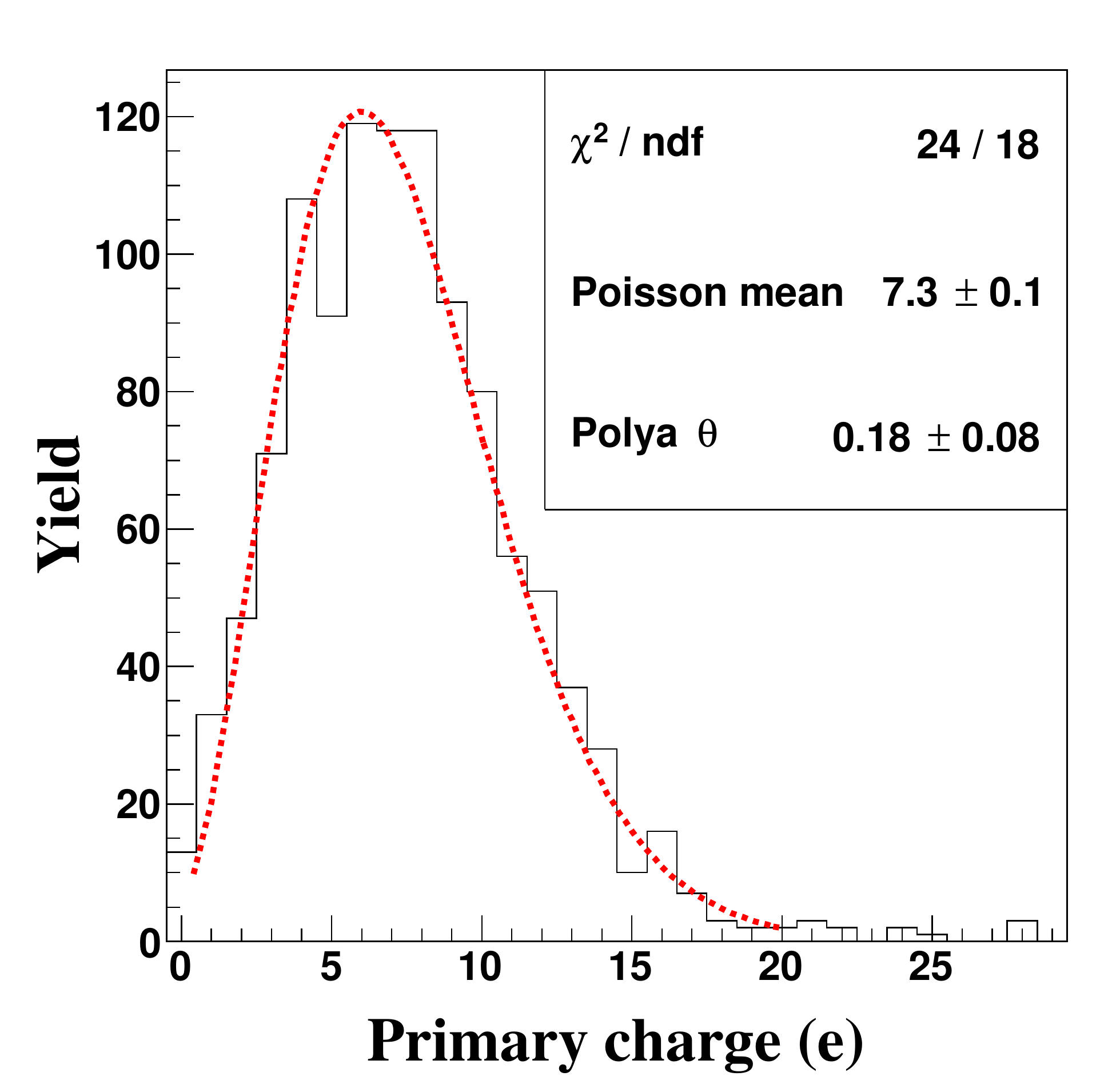}
   	\caption{\label{fig:response-e} {Charge distribution of the prototype for electrons. The dotted line shows
a Poisson distribution convoluted with Polya and Gaussian distributions~\cite{e16-hbd2}. } }
 \end{center}
\end{figure}

The performance of the prototype was evaluated at the J-PARC K1.1BR beam line using a negatively charged secondary beam at 1.0~GeV/$c$ containing about 20\% electrons.
Two scintillators were placed in front and behind the prototype to trigger on beam particles within an area of 10 $\times$ 10 mm${}^2$, and two gas Cherenkov detectors were placed in series to identify the beam particles independently.
During the beam test, the water and oxygen contents of the gas were maintained below 5~ppm and 0.5~ppm, respectively, which resulted in a transparency greater than 99\%.

Figure~\ref{fig:response-e} shows the charge distribution obtained by the prototype for electrons.
The spectrum was fitted to a Poisson distribution convoluted with Polya and Gaussian distributions.
The number of photoelectrons (the Poisson mean) was only $7.3 \pm 0.1$ because 
the hole pitch of the top GEM was 240~$\mu$m.
Such a wide pitch was intended to make the effective sensitive area larger to 
increase the number of photoelectrons. However, this resulted in a significant
loss of photoelectrons due to its very small transport efficiency ($17 \pm 10 \%$).
The larger hole pitch requires that the photoelectrons must travel a longer distance to reach the GEM holes and therefore have a higher probability of attachment and scattering by the gas.
In the E16 experiment, 55/110-GEMs will be used which will have a much higher transport efficiency.

\begin{figure}[htb]
 \begin{center}
   \includegraphics[width=6cm] {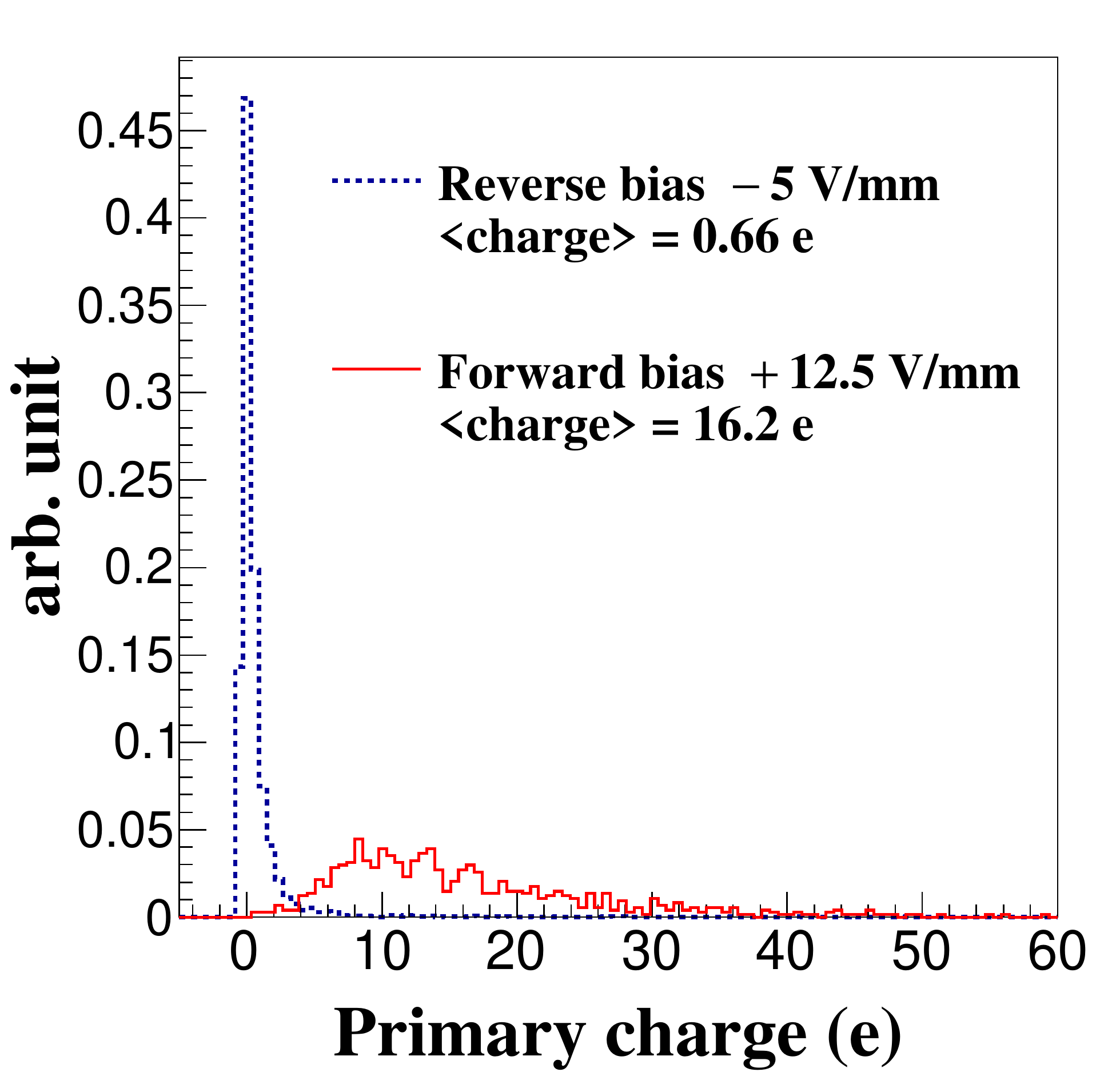}
   	\caption{\label{fig:response-pi} {Charge distribution of the prototype for pions
     with a forward bias of 12.5~V/mm (red solid) and a reverse bias of -5~V/mm (blue dashed)~\cite{e16-hbd2}.} }
 \end{center}
\end{figure}

Figure~\ref{fig:response-pi} shows the charge distribution of the prototype for pions. A reverse bias of 5~V/mm efficiently suppresses the signal for pions.
The residual signal is primarily due to the ionization produced very close to the top GEM surface and inside the first transfer gap.

\begin{figure}[ht]
 \begin{center}
   \includegraphics[width=7.5cm] {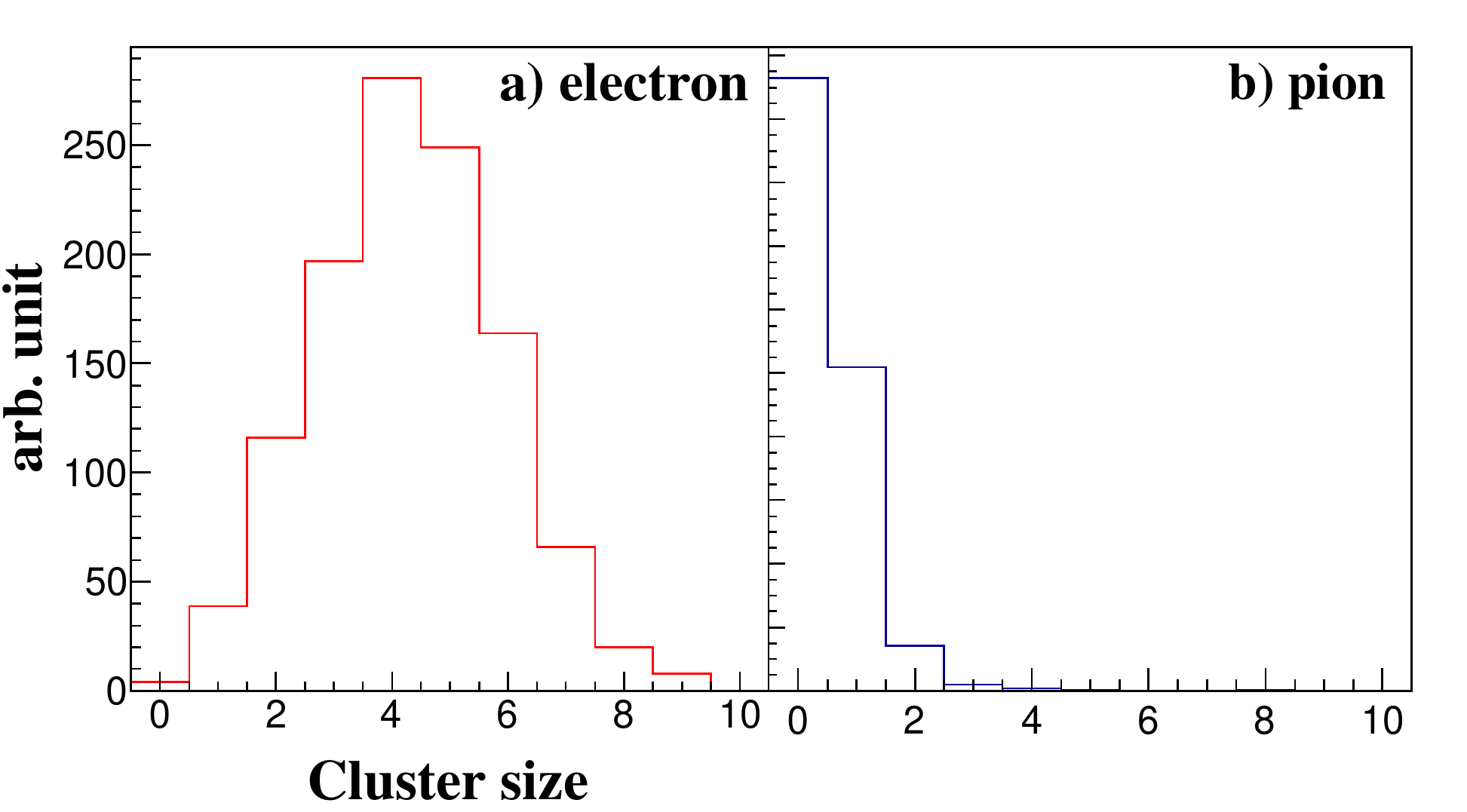}
   	\caption{\label{fig:cluster-size} {Cluster size distributions for a) electrons and b) pions~\cite{e16-hbd2}.} }
 \end{center}
\end{figure}

The Cherenkov image on the photocathode is 34~mm in diameter while
the ionization loss is localized to a few mm.
Therefore, their signal distributions across neighboring pads are very different and can be utilized to reject pions. 
A pad is considered to be hit if the ADC value is larger than its pedestal by 7$\sigma$.
This threshold corresponds to 0.22 photoelectrons. A cluster is defined as the number of neighboring hit pads.
Two pads are considered as neighbors when they share at least one corner in common.
Figure~\ref{fig:cluster-size} displays cluster size distributions for electrons and pions, which clearly demonstrate the discrimination power of using the cluster size. 

 \begin{figure}[ht]
 \begin{center}
   \includegraphics[width=7.5cm] {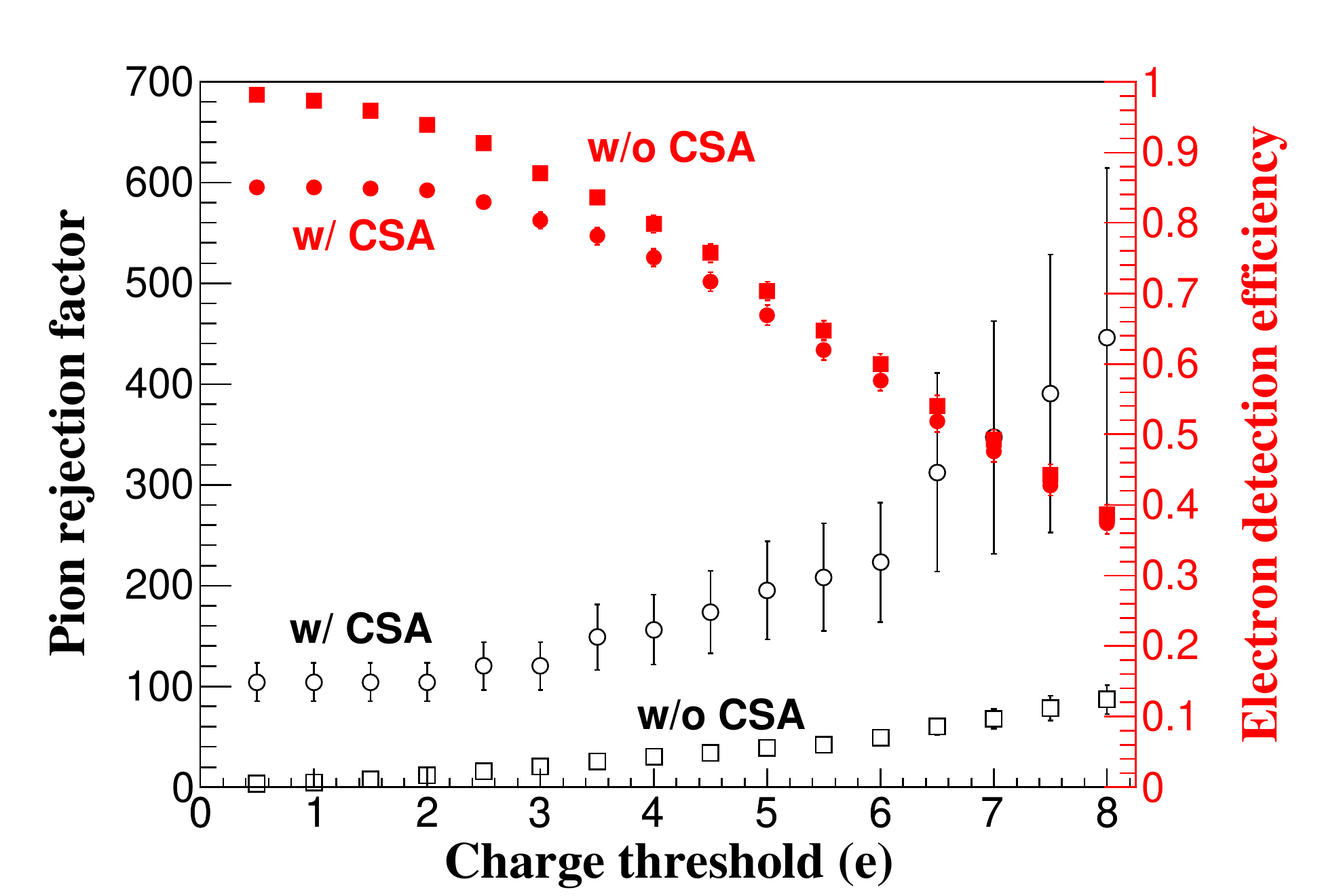}
   	\caption{\label{fig:rej-eff} {Rejection factor of pions (open symbols) and efficiency of electrons (red filled symbols) as a function of 
   charge threshold with and without the cluster size analysis (circle and box, respectively)~\cite{e16-hbd2}.} }
 \end{center}
\end{figure}

Figure~\ref{fig:rej-eff} shows the rejection factor for pions and the efficiency for electrons.
With the help of the cluster size analysis, a pion rejection factor of 120 with electron efficiency of 84\% was obtained 
when a charge of more than 2.5 photoelectrons and a cluster size of three or more were required.
Without the cluster size analysis, the pion rejection factor was only 16 with an electron efficiency of 91\%
at the same threshold for photoelectrons.
A cut on cluster size is therefore very powerful in effectively rejecting pions while maintaining a reasonable efficiency for electrons.


 \section{Future Applications}

\begin{figure}[th]
 \begin{center}
   \includegraphics[width=7.2cm] {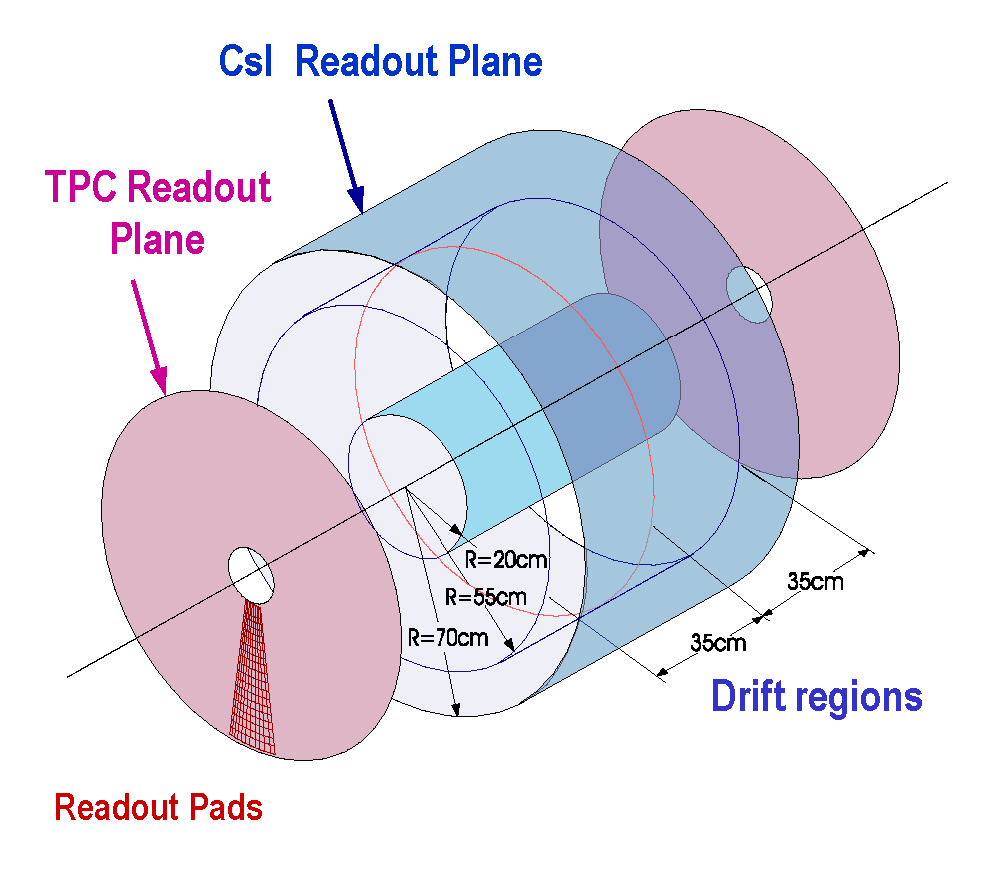}
   	\caption{\label{fig:TPC_HBD_Concept} {Conceptual design of the TPC-HBD detector originally proposed for PHENIX.} }
 \end{center}
\end{figure}

Identification of electrons and hadrons is often a part of the overall particle identification strategy in many experiments and there are therefore potentially many future applications for Hadron Blind Detectors in nuclear and particle physics. Since an HBD is essentially a proximity focused Cherenkov detector, it can be used in any application that requires such a detector. In addition, an HBD can in principle also be combined with a tracking detector, using the radiator volume as the drift volume of a TPC, thus providing particle id and tracking information in the same detector. This idea was originally proposed during the early stages of considering an HBD for the PHENIX experiment. Figure ~\ref{fig:TPC_HBD_Concept} shows a conceptual design of a TPC/HBD detector that was originally proposed for PHENIX. The detector was never implemented due to the added cost and complexity of such a detector, but a  prototype study was later carried out to demonstrate a proof of the concept \cite{Azmoun-2019}. A similar design, which would use an HBD combined with a radial drift TPC, is currently being considered as a possible detector at the future Electron Ion Collider. Both of these applications are discussed below.


\subsection{TPC/HBD Proof of Concept}

\begin{figure}[th]
 \begin{center}
   \includegraphics[width=7.5cm] {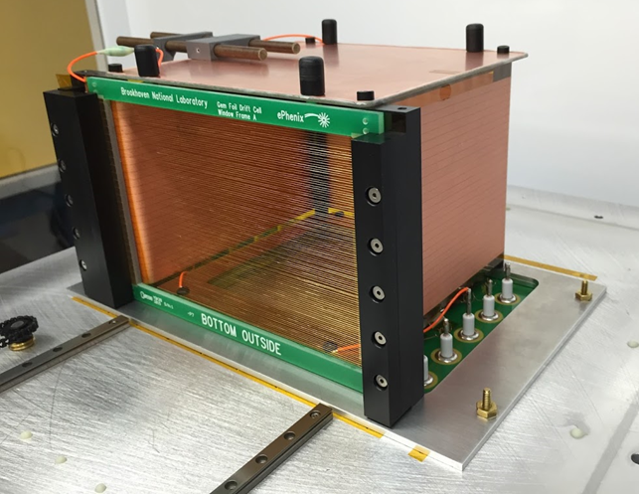}
   	\caption{\label{fig:TPCC_Field_Cage} {Field cage of the TPCC detector with a wire plane on one side to allow passage of UV light to the Cherenkov detector \cite{Azmoun-2019}.} }
 \end{center}
\end{figure}

   A small prototype TPC-Cherenkov detector, or TPCC, was built and tested in order to demonstrate the TPC/HBD concept. The prototype was tested in the test beam at Fermilab and the results are now published \cite{Azmoun-2019}. Figure ~\ref{fig:TPCC_Field_Cage} shows the field cage of the TPCC detector. Three sides consisted of a kapton foil with copper strips that created a drift field in the vertical direction, and one side consisted of thin wires that were biased to the same potentials as the strips that allowed the passage of UV light to the Cherenkov detector. The field cage, which measured approximately $10 \times 10 \times 10$ cm$^3$, created the drift volume for the TPC as well as the radiator for the Cherenkov detector. Charge produced in the drift region drifted to the bottom of the detector and was measured using a quadruple GEM detector with a chevron pad readout. 
  
\begin{figure}[th]
 \begin{center}
   \includegraphics[width=7.5cm] {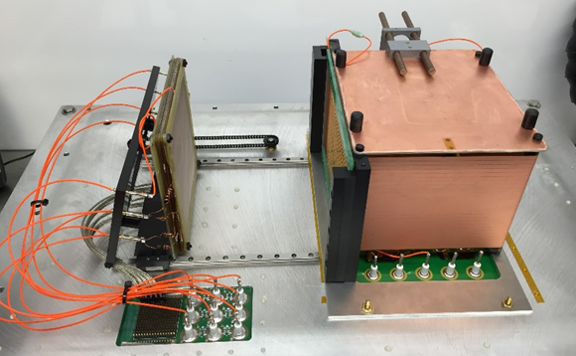}
   	\caption{\label{fig:TPCC_Interior_View} {Interior view of the TPCC detector showing the TPC detector on the right and the Cherenkov detector on its movable rail on the left \cite{Azmoun-2019}.} }
 \end{center}
\end{figure}

  The Cherenkov detector consisted of another quadruple GEM detector that was equipped with a CsI photocathode on the top surface of the top GEM and was mounted on a set of rails just outside the drift volume beyond the wire plane. The readout plane for the Cherenkov GEM was only coarsely segmented into a $3 \times 3$ array of 3.2 cm square pads that allowed detection of the Cherenkov light but provided very little position information. Figure ~\ref{fig:TPCC_Interior_View} shows the arrangement of the two detectors with the TPC on the right and the Cherenkov detector on the left. Both detectors were enclosed in a gas tight enclosure that was filled with CF$_4$ that served as the radiator gas for the Cherenkov detector, the drift gas for the TPC and the operating gas for the GEMs.

\begin{figure}[th]
 \begin{center}
   \includegraphics[width=7.5cm] {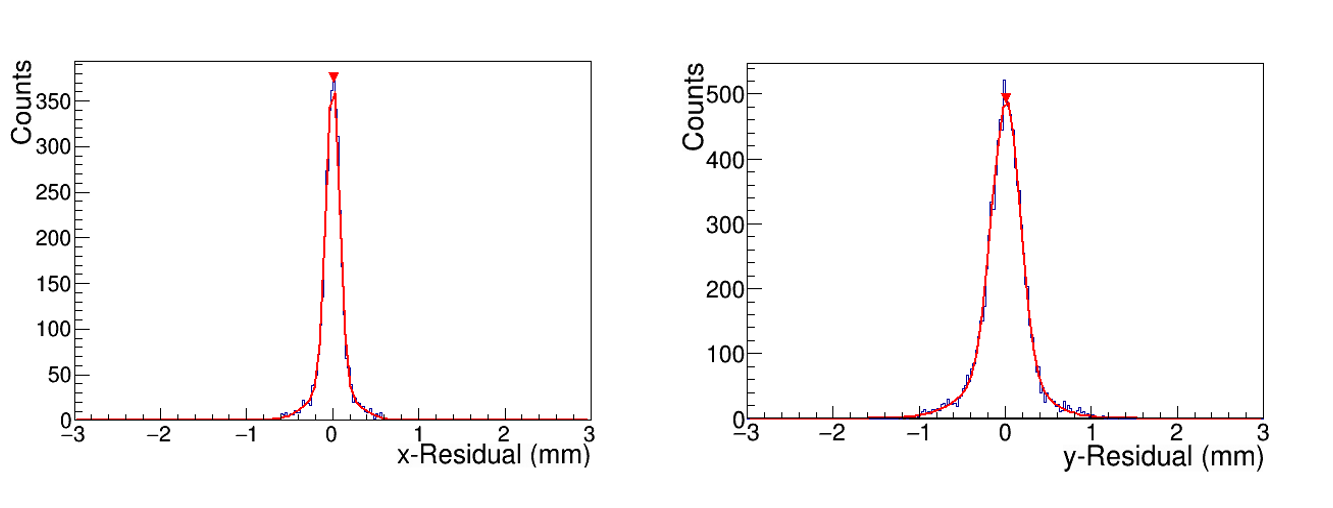}
   	\caption{\label{fig:TPCC_Residual_Distributions} {Track resolutions measured in the TPC portion of the detector giving 80 $\mu$m in X and 167 $\mu$m in Y \cite{Azmoun-2019}.} }
 \end{center}
\end{figure}

  The detector was tested in the test beam at Fermilab in order to demonstrate a proof of the TPC/HBD concept. Tracks were reconstructed in the TPC portion of the detector and compared to the same tracks measured in a high resolution silicon telescope in the beam. Figure ~\ref{fig:TPCC_Residual_Distributions} shows the residual distribution for tracks measured in the TPCC detector which gave a resolution of 80 $\mu$m in the X coordinate  and 167 $\mu$m and in the Y coordinate.

\begin{figure}[th]
 \begin{center}
   \includegraphics[width=7.5cm] 
   {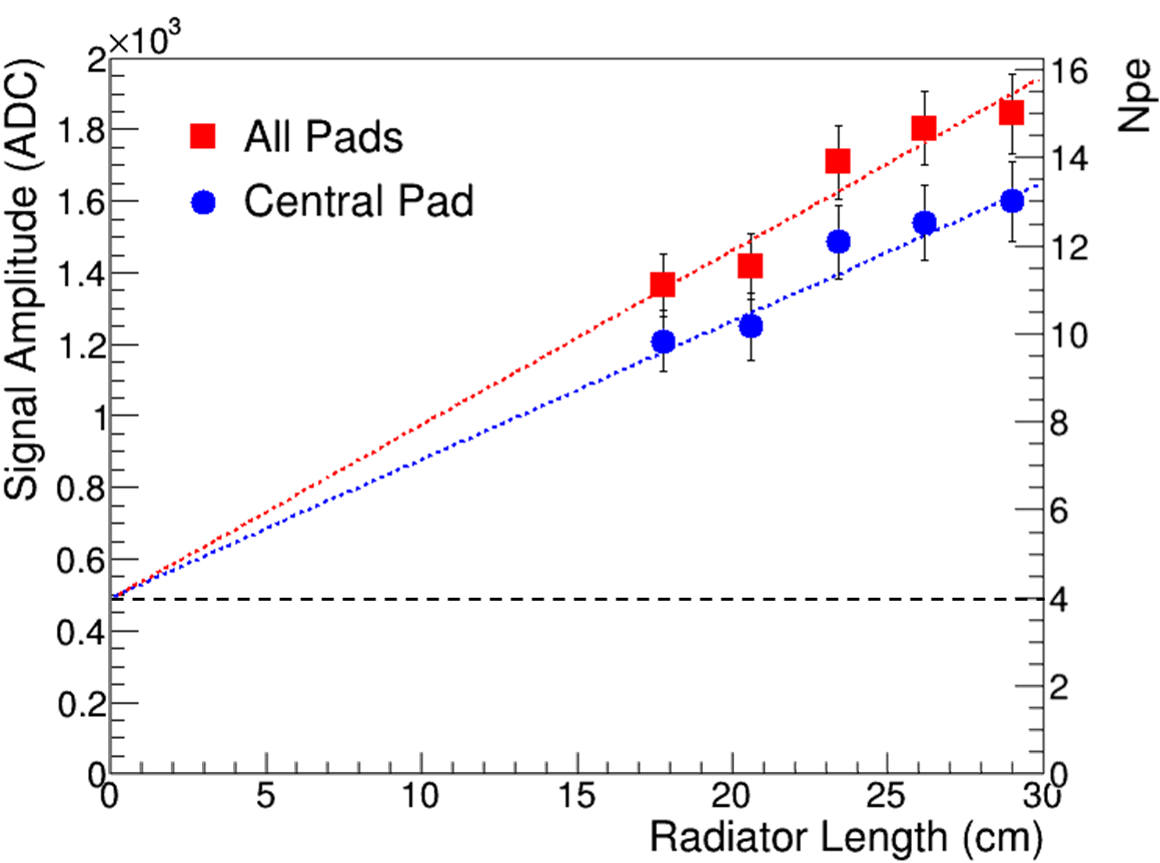}
   	\caption{\label{fig:TPCC_Photoelectron_Yield} {Photoelectron yield (right axis) measured by the Cherenkov detector as a function of the radiator length. Blue (round) points are for the central pad only and red (square) points are for all pads \cite{Azmoun-2019}.} }
 \end{center}
\end{figure}

   The Cherenkov portion of the TPCC was tested using a mixed beam of electrons and hadrons. Particles were identified using a differential Cherenkov in the beam line and electrons were used to measure the photoelectron yield. The yield was measured as a function of the radiator length by moving the Cherenkov detector along its rails and is shown in Fig.~\ref{fig:TPCC_Photoelectron_Yield}. The overall yield for the sum of all pads hit for a radiator length of 29 cm was $\sim$ 15 p.e., which contained a contribution of $\sim$ 4 p.e. from the residual hadron signal. After subtracting the hadron background, the photoelectron yield for Cherenkov light was $\sim$ 11 p.e., which was consistent with the yield measured in the PHENIX HBD. 

\begin{figure}[th]
 \begin{center}
   \includegraphics[width=7.5cm] {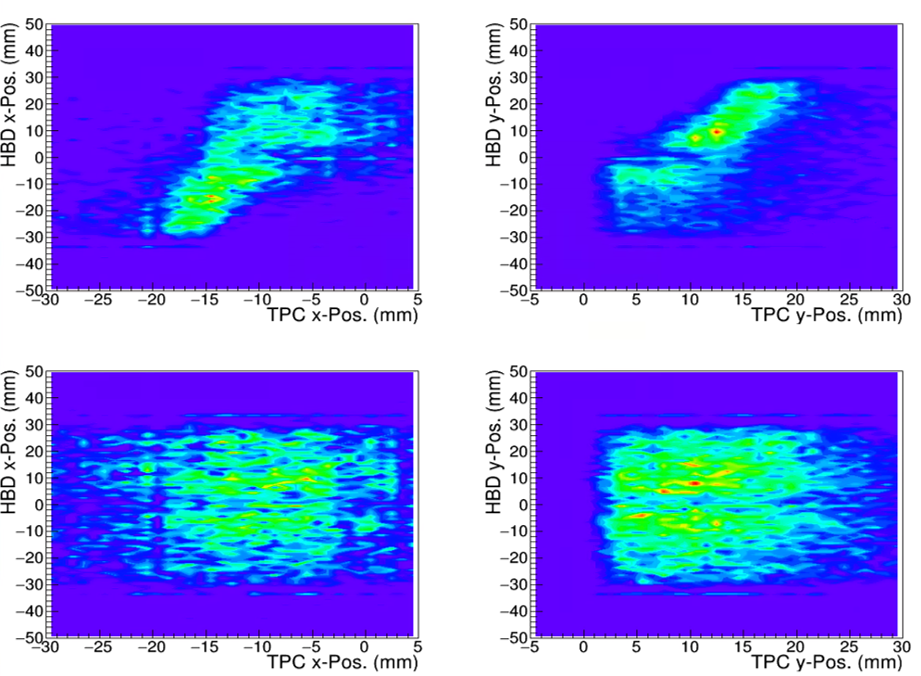}
   	\caption{\label{fig:TPC_Cherenkov_Correlations} {Correlations between the projection of the track found in the TPC with the blob measured in the Cherenkov detector. Top shows correlations in X and Y. Bottom shows the correlation of the blob position with randomized hits from the TPC. The spatial correlation was limited by the coarse pad size in the Cherenkov detector \cite{Azmoun-2019}.} }
 \end{center}
\end{figure}

  Due to the rather coarse segmentation of the readout pads in the Cherenkov detector it was not possible to obtain a precise measurement of the correlation between the track found in the TPC and the position of the blob measured with the Cherenkov. Nevertheless, a correlation was measured which showed that the Cherenkov blobs were indeed correlated with the TPC tracks. Figure ~\ref{fig:TPC_Cherenkov_Correlations} shows this correlation in the X and Y coordinates of Cherenkov detector compared with the correlation between the blobs with randomized hits from the TPC tracks. While the correlation would have been much clearer with finer segmentation of the readout plane, it was sufficient to demonstrate the concept. 

\subsection{EIC Applications}

  The Electron Ion Collider (EIC) is a new accelerator that has been proposed that would collide a beam of electrons with beams of protons or ions to study nuclear structure and the spin properties of nucleons with much greater precision than has been achieved in the past \cite{EIC-White-Paper-2014} .
  The project received CD-0 approval by the US Department of Energy in January 2020 and will be sited at Brookhaven National Lab. The total project cost will be $\sim$ 2B US$\$$ and will become the largest nuclear physics facility in the US.   
  
  Measurements at the EIC will involve measuring deep inelastic scattering of electrons off the partons inside the nucleons and would require a precise measurement of the energy and/or momentum and direction of the scattered electron, as well as identifying the electron in a background of produced hadrons. A Hadron Blind Detector, or a TPC/HBD, can provide a means of achieving both types of measurements.

\begin{figure}[th]
 \begin{center}
   \includegraphics[width=7.5cm] {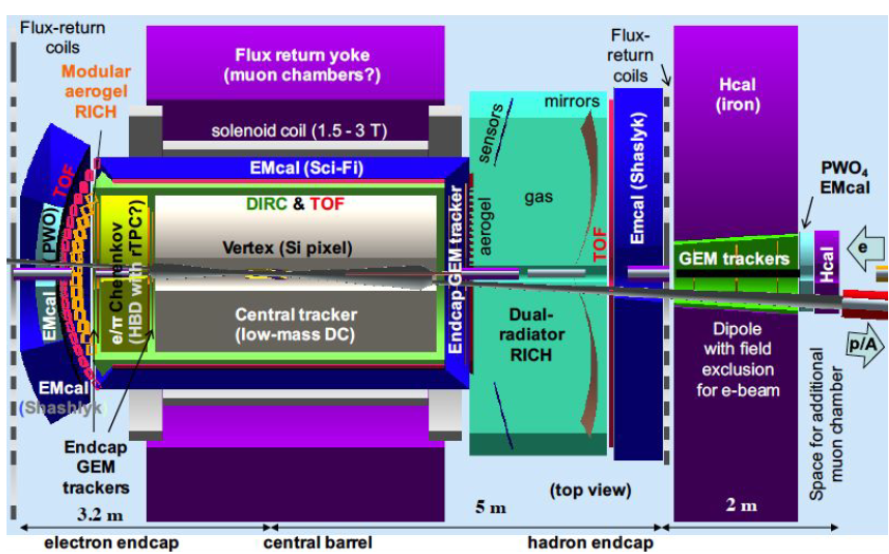}
   	\caption{\label{fig:JLEIC_Detector} {Conceptual design for an EIC detector developed at JLAB. A TPC-HBD detector would be used in the electron going direction to help identify the scattered electron and reject hadron background. The Dual Radiator RICH on the hadron going side could also utilize an HBD like detector for identifying high momentum hadrons  \cite{EIC-White-Paper-2014}.} }
 \end{center}
\end{figure}

 One such scheme is shown in Fig. ~\ref{fig:JLEIC_Detector}, which shows a conceptual design for an EIC detector that was developed at JLAB.  A TPC/HBD detector would be located in the forward region of the electron-going direction where the electron would be scattered at relatively small angles. Due to the solenoidal magnetic field in the proposed design, the electron track undergoes very little bending at small angles, so in order to optimize the momentum measurement, the drift of the TPC would be in the radial direction. The TPC readout would be around the perimeter of the detector while the Cherenkov readout would be at the exit face. The hadron beam enters the detector from the opposite direction and secondary hadrons that are scattered backward into the electron endcap detector would be identified using the TPC/HBD in conjunction with the planar GEM tracking detectors and the endcap electromagnetic calorimeter, or possibly an aerogel Cherenkov counter. 
 
  On the hadron-going side, there is also a Dual Radiator RICH for identifying produced hadrons over a large momentum range. It will consist of a radiator for low momentum particles (e.g., aerogel) along with a gas radiator for high momentum particles. In order to identify high momentum hadrons (up to 50 GeV/c) the RICH must be able to separate rings produced by pions, kaons and protons with very high precision. One such scheme that combined the features of the PHENIX and E16 HBDs with a detector configuration similar to the one used by CERES was studied by the group at Stony Brook University \cite{GEM-RICH-EIC-2015}. It utilized a 1 m long CF$_4$ radiator and a mirror to reflect the Cherenkov photons back to a UV detector consisting of a 5 stage GEM detector with a CsI photocathode and a readout consisting of 5 mm hexagonal pads. It achieved very good $\pi$/K/p separation at 20, 25 and 32 GeV/c, and calculations based on their measurements showed that one can obtain up to $\sim 3 \sigma$ separation for momenta up to 60 GeV/c using smaller pads. While this detector is not strictly a hadron blind detector, it does utilize many of the features of an HBD that have been implemented along the course of its development.

 \section{Summary}
\label{}

The concept of a Hadron Blind Detector was originally developed in the early 1990's based on the need to identify highly relativistic leptons (mainly electrons) within a large background of accompanying hadrons. This requirement occurs in a number of different physics measurements, such as identifying heavy quark decays, lepton identification within jets and identifying lepton pairs in heavy-ion collisions within a large background of neutral meson decays. The concept involves the combination of a gaseous radiator and a deep UV sensitive photocathode such as CsI in a windowless configuration in order to achieve a high photodetection yield and efficiency. Relativistic particles produce Cherenkov light in the radiator while hadrons which are below the Cherenkov threshold do not. The first implementation of an HBD used Ring Imaging Cherenkov Detectors, originally proposed by Ypsilantis, in the CERES experiment at CERN. This was later implemented in a proximity focused (non ring imaging) configuration in the PHENIX experiment at RHIC and again in a similar HBD for experiment E16 at J-PARC. All three of these experiments used the HBD to measure low-mass electron pairs in pA and AA collisions. An extension of the HBD concept was proposed to combine the features of an HBD with a TPC in order to provide both tracking and particle ID in the same detector. A proof of principle of this concept was demonstrated in a small prototype detector and shown to work, and a version of a combination TPC/HBD has been proposed in an experiment for a future Electron Ion Collider.

\section{Acknowledgements}
\label{}

The authors wish to acknowledge the support of their home institutions in the preparation of this article. Brookhaven National Laboratory is supported by the US Department of Energy, Division of Nuclear Physics, under Contract DE-SC0012704. 

\nocite{*}
\bibliographystyle{elsarticle-num}
\bibliography{jos}




 



\end{document}